\documentclass[UKenglish,twocolumn,epjc3]{svjour3}          

\usepackage[a4paper, left=1.3cm, right=1.3cm, top=2.5cm, bottom=2cm]{geometry}

\usepackage{placeins}
\usepackage{relsize}

\usepackage{mathtools}
\usepackage{lineno}

\usepackage{cite}

\usepackage{atlaspackage}

\usepackage{atlasphysics}

\graphicspath{{../logos/}{figures/}}

\hypersetup{pdftitle={ATLAS draft},pdfauthor={The ATLAS Collaboration}}
\institute{CERN, CH-1211 Geneva 23. Switzerland}

\newcommand{\zjpsi}{\ensuremath{Z\, +\, \Jpsi}\xspace}
\newcommand{\zpjpsi}{{\ensuremath{Z\, +}} prompt \ensuremath{\,\Jpsi}\xspace}
\newcommand{\znpjpsi}{{\ensuremath{Z\, +}} non-prompt \ensuremath{\,\Jpsi}\xspace}

\usepackage{bigdelim,multirow} 

\journalname{Eur. Phys. J. C}

\usepackage[font=small,labelfont=bf,labelsep=space]{caption}
\usepackage{graphicx}
\usepackage{preprintcover}

\PreprintCoverPaperTitle{Observation and measurements of the production of prompt and non-prompt {\boldmath $J/\psi$} mesons in association with a {\boldmath $Z$} boson in {\boldmath $pp$} collisions at {\boldmath $\sqrt{s}= 8\,\mathrm{TeV}$} with the ATLAS detector}  

\PreprintIdNumber{CERN-PH-EP-2014-276}  

\PreprintCoverAbstract{The production of a $Z$ boson in association with a $J/\psi$ meson in proton--proton collisions probes the production mechanisms of quarkonium and heavy flavour in association with vector bosons, and allows studies of multiple parton scattering. Using $20.3\,\mathrm{fb^{-1}}$ of data collected with the ATLAS experiment at the LHC, in $pp$ collisions at $\sqrt{s}=8\,\mathrm{TeV}$, the first measurement of associated $Z\, +\, J/\psi$ production is presented for both prompt and non-prompt $J/\psi$ production, with both signatures having a significance in excess of $5\sigma$. The inclusive production cross-sections for $Z$ boson production (in $\mu^+\mu^-$ or $e^+e^-$ decay modes) in association with prompt and non-prompt $J/\psi(\to\mu^+\mu^-)$ are measured relative to the inclusive production rate of $Z$ bosons in the same fiducial volume to be $(36.8\pm 6.7\pm 2.5)\, \times\, 10^{-7}$ and $(65.8\pm 9.2\pm 4.2)\, \times\, 10^{-7}$ respectively. Normalised differential production cross-section ratios are also determined as a function of the $J/\psi$\ transverse momentum. The fraction of signal events arising from single and double parton scattering is estimated, and a lower limit of $5.3\ (3.7)\,\mathrm{mb}$ at $68\ (95) \%$ confidence level is placed on the effective cross-section regulating double parton interactions.
}

\PreprintJournalName{Eur. Phys. J. C}

\begin{document}
\sloppy

\title{Observation and measurements of the production of prompt and non-prompt {\boldmath $\Jpsi$} mesons in association with a {\boldmath $Z$} boson in {\boldmath $pp$} collisions at {\boldmath $\sqrt{s}= 8\,\mathrm{TeV}$} with the ATLAS detector}
\titlerunning{Production of prompt and non-prompt \Jpsi\ mesons in association with a $Z$ boson.}

\author{The ATLAS Collaboration}

\date{Received: date / Accepted: date}

\maketitle

\begin{abstract}
The production of a $Z$ boson in association with a $J/\psi$ meson in proton--proton collisions probes the production mechanisms of quarkonium and heavy flavour in association with vector bosons, and allows studies of multiple parton scattering. Using $20.3\,\mathrm{fb^{-1}}$ of data collected with the ATLAS experiment at the LHC in $pp$ collisions at $\sqrt{s}=8\,\mathrm{TeV}$, the first measurement of associated $Z\, +\, J/\psi$ production is presented for both prompt and non-prompt $J/\psi$ production, with both signatures having a significance in excess of $5\sigma$. The inclusive production cross-sections for $Z$ boson production (analysed in $\mu^+\mu^-$ or $e^+e^-$ decay modes) in association with prompt and non-prompt $J/\psi(\to\mu^+\mu^-)$ are measured relative to the inclusive production rate of $Z$ bosons in the same fiducial volume to be $(36.8\pm 6.7\pm 2.5)\, \times\, 10^{-7}$ and $(65.8\pm 9.2\pm 4.2)\, \times\, 10^{-7}$ respectively. Normalised differential production cross-section ratios are also determined as a function of the \Jpsi\ transverse momentum. The fraction of signal events arising from single and double parton scattering is estimated, and a lower limit of $5.3\ (3.7)\,\mathrm{mb}$ at $68\ (95) \%$ confidence level is placed on the effective cross-section regulating double parton interactions.
\end{abstract}


\section{Introduction}

In the Standard Model, a single parton--parton interaction can produce a $J/\psi$ meson in association with a $Z$ boson
either through a ``prompt''  QCD subprocess, or through the production of a $Z$ boson with a
$b$-quark and its subsequent decay into a $J/\psi$ (``non-prompt'' production). The same classification into prompt/non-prompt applies to any feed-down into \Jpsi production from the decays of excited charmonium states (expected to be approximately $20-30\%$ of the total inclusive rate), depending on the production mechanism for those states. In addition, this final state may also result from the production of a $Z$ boson and a $J/\psi$ (either promptly or non-promptly produced) from two distinct parton--parton interactions within the same proton--proton collision. Previous searches for the related processes $W\,+\,\Upsilon(1S)$ and $Z\,+\,\Upsilon(1S)$ by CDF saw no evidence for the associated-production of vector-bosons and quarkonia and set limits on the production rate~\cite{Acosta:2003mu,Aaltonen:2014rda}. The production of a prompt \Jpsi in association with a $W$ boson was observed previously~\cite{Aad:2014rua} by the ATLAS experiment.

The mechanisms responsible for the production of prompt $J/\psi$, and
indeed all quarkonia, are not fully understood in hadron
collisions. While the rate of hadroproduction of the
\Jpsi~\cite{JpsiATLAS7TeV,Aaij:2011jh,Abelev:2012gx,Chatrchyan:2011kc}
and
$\psi(2S)$ ~\cite{Chatrchyan:2011kc,ATLASpsi2sjpsipipi,Aaij:2012ag,Abelev:2014qha}, as a function of their transverse momentum, $p_\mathrm{T}$, is now modelled well by predictions in the non-relativistic QCD (NRQCD)~\cite{Caswell:1985ui,Bodwin:1994jh,NRQCD1} framework up to transverse momenta of 100~GeV, predictions of related observables such as charmonium spin-alignment~\cite{Brambilla:2010cs,Brambilla:2004wf} remain challenging to model simultaneously with the production rate, in part due to the number of free parameters which are not calculable and must be constrained from data. The study of additional observables and new final states provides further constraints on the 
contributions from colour-singlet~\cite{CSM1,CSM2,CSM3,CSM4,CSM5,CSM6,CSM7} and colour-octet production processes, and their properties.
The production of a gauge boson in association with a \Jpsi sets a high energy scale for the scattering process and results in an improvement in the perturbative convergence of the calculations~\cite{Gong,ass2} that has troubled the accuracy of quarkonium production models in the past~\cite{Lansberg:2008gk}. Recent literature~\cite{ass2} has suggested that colour-octet contributions should dominate the total production rate and that next-to-leading-order (NLO) contributions enhance the cross-section over leading-order (LO) predictions, while other groups~\cite{Gong} state that colour-singlet processes may be important.

Contributions to the total $Z\,+\,J/\psi$ production rate can come from non-prompt $J/\psi$ originating from the decay of a $b$-hadron. Measurement of this contribution provides a new opportunity for studying heavy--flavour production in association with a $Z$ boson~\cite{Aad:2011jn,Chatrchyan:2014dha}.
Beyond the study of quarkonium production mechanisms, measurement of the $Z\,+$ prompt $J/\psi$ rate may be relevant for the study of $ZZ^*$ production in a kinematic regime complementary to that previously studied at the Large Hadron Collider (LHC)~\cite{Aad:2012awa,Khachatryan:2014dia} where one on-shell $Z$ boson is produced along with a highly virtual boson that fragments into a $c\overline{c}$ pair. 
Measurement of \zpjpsi production also represents an important background to the search for the rare $Z\to\ell^+\ell^-\Jpsi$ three-body decay \cite{Zpjill1,Zpjill2,Zpjill3}.
In the future, $Z\,+$ prompt $J/\psi$ may prove to be a compelling mode for the study of rare decays of the Higgs boson in quarkonia and associated vector-boson decay modes, proposed in Refs.~\cite{Doroshenko:1987nj,Kartvelishvili:1988pu}
 and more recently in Refs.~\cite{Higgs2,Higgs3}. Such decays have received renewed attention as a promising mode for the study of Higgs boson charm couplings~\cite{Bodwin:2013gca} and its CP properties~\cite{Bhattacharya:2014rra},  and also as a possible background to $H\to ZZ^*$ decay~\cite{Gao:2014xlv}.  The production of a $Z$ boson in association with a $J/\psi$ can also contribute to the search for new physics~\cite{Davoudiasl:2012ag,Curtin:2013fra,Falkowski:2014ffa,Higgs2,Clarke:2013aya}.

In addition to the production of \zjpsi via single parton scattering (SPS) processes, double parton scattering (DPS) interactions~\cite{Paver:1982yp,Sjostrand:2004pf,Korotkikh:2004bz,Gaunt:2009re,Bartalini:2011jp,Golec-Biernat:2014nsa,Gaunt:2014rua} are expected to constitute a significant proportion of the observed signal. While DPS processes are not distinguishable event-by-event from SPS interactions, azimuthal angular correlations between the $Z$ and the \Jpsi are expected to be starkly different for the two processes, allowing information on their relative contributions to be extracted. These data can be used to tune the modelling of multiple interactions in other high-energy hadron--hadron processes.

This paper presents a measurement of the cross-section for the associated-production of $Z$ and $J/\psi$ relative to inclusive $Z$ production. The results are shown as {\em fiducial} cross-section ratios defined in a restricted phase-space of the muons from $J/\psi$ decay, and also as {\em inclusive} cross-section ratios after correcting for the $J/\psi$ kinematic acceptance of these muons, for the range of $J/\psi$ transverse momentum $8.5-100\,\mathrm{GeV}$ and rapidity $|y(J/\psi)|<2.1$. The contributions from prompt and non-prompt $J/\psi$ production are presented separately. The cross-section ratio for single parton scattering is obtained after estimating and subtracting the contribution due to double parton scattering. A lower limit on the effective cross-section regulating double parton interactions is presented. Differential cross-section ratios as a function of the transverse momentum $p_\mathrm{T}$ of the $J/\psi$ are shown for prompt and non-prompt production, inclusive and DPS modes.

\section{The ATLAS detector}\label{sec:atlas_detector}
The ATLAS detector~\cite{ATLASdetector} is a general-purpose detector with a cylindrical
geometry\footnote{ATLAS uses a right-handed coordinate system with its origin at the nominal interaction point (IP) in the centre of the detector and the $z$-axis along the beam pipe. 
The $x$-axis points from the IP to the centre of the LHC ring, and the $y$-axis points upward. Cylindrical coordinates $(r,\phi)$ are used in the transverse plane, 
$\phi$ being the azimuthal angle around the beam pipe. The pseudorapidity $\eta$ is defined in terms of the polar angle $\theta$ as $\eta=-\ln\tan(\theta/2)$ and the 
transverse momentum $p_{\rm T}$ is defined as $p_{\rm T}=p\sin\theta$. The rapidity is defined as $y=0.5\ln\left(\left( E + p_z \right)/ \left( E - p_z \right)\right)$, 
where $E$ and $p_z$ refer to energy and longitudinal momentum, respectively. The $\eta$--$\phi$ distance between two particles is defined as 
 $\Delta R=\sqrt{(\Delta\eta)^2 + (\Delta\phi)^2}$.} and forward-backward symmetric coverage in pseudorapidity $\eta$. 
The detector consists of inner tracking detectors, calorimeters and a muon spectrometer, and has a three-level trigger system.
The inner tracking detector (ID) is composed of a silicon pixel detector, a semiconductor microstrip detector (SCT) and a transition radiation tracker (TRT).
The ID directly surrounds the beam pipe and is immersed in a 2\,T axial magnetic field generated by a superconducting solenoid. 

The calorimeter system surrounds the solenoid and consists of a highly granular liquid-argon electromagnetic calorimeter (EM) and a steel/scintillator tile hadronic calorimeter. The EM calorimeter has three layers: the first consists of fine-grained strips in the $\eta$ direction, the second collects most of the energy deposited in the calorimeter by photon and electron showers, and the third provides measurements of energy deposited in the tails of these showers. Two complementary presampler detectors complete the EM, correcting for energy lost in the material before the calorimeter. This fine segmentation provides electron identification in conjunction with the inner detector in the region $|\eta|<2.5$.

The muon spectrometer (MS) surrounds the calorimeters and consists of three large air-core superconducting magnets (each with eight coils), which generate a toroidal magnetic field. The MS is instrumented in three 
layers with detectors (monitored drift tubes and cathode strip chambers) that provide precision muon tracking covering $|\eta| < 2.7$ and fast trigger detectors (resistive plate chambers 
and thin gap chambers) covering the range $|\eta| < 2.4$.

The ATLAS trigger is a three-level system~\cite{ATLAS:trig} (Level-1, Level-2 and Event Filter) used to reduce the $20\,\mathrm{MHz}$ proton bunch collision rate to a several-hundred Hz event transfer rate recorded to mass storage.
The system consists of a Level-1 trigger implemented in hardware and a software-based two-stage High Level Trigger (HLT). 
The Level-1 system provides a rough measurement of lepton candidate position in ``regions of interest'' (RoI) with a spatial granularity of $\Delta\varphi\times\Delta\eta\approx 0.1 \times 0.1$. 
These RoI are used to seed HLT algorithms that use higher precision MS, ID and EM measurements to reconstruct lepton trigger objects.

\section{Event selection and reconstruction} \label{sec:eventselection}

Events are collected by triggers requiring at least one lepton with $p_\mathrm{T} > 24\,\mathrm{GeV}$. These triggers are highly efficient in collecting $Z\to\ell^+\ell^-$ decays and were not prescaled during the 2012 data-taking period. 
Triggered events are required to satisfy certain standardised data-quality requirements, which exclude events taken when temporary faults in detector systems compromise the reconstruction. The total integrated luminosity of proton--proton collisions used in this measurement, after data-quality requirements are applied, is $20.3\,\mathrm{fb^{-1}}$. 

The final state of this measurement is $Z(\to \ell^+\ell^-)+J/\psi(\to\mu^+\mu^-)$, where $\ell=\mu,e$, and therefore candidate events are required to have two pairs of leptons with opposite charge. 
Each pair of leptons is then fitted to a common vertex, with the invariant mass of the first pair required to be close to the  $Z$ boson mass and that of the second pair to be near the $J/\psi$ mass. For events with more than four leptons, all possible combinations of $\ell^+\ell^-$ and $\mu^+\mu^-$ pairs are considered. In rare cases where ambiguous solutions are found, the pairings giving the dilepton combination with mass closest to the particle ($Z$ or $J/\psi$) world-average mass are chosen.

\subsection{Lepton reconstruction}
Muons are identified~\cite{muons} by tracks (or track segments) reconstructed in the MS, matched to tracks reconstructed in the ID. Track reconstruction in the inner detector uses the measurements from the pixel, SCT and TRT detectors. The ``inside-out'' reconstruction strategy starts by finding a track candidate in the pixel and SCT detectors and then extends the trajectories of successfully fitted tracks to the TRT to reconstruct a full inner detector track. Outside of the TRT acceptance ($|\eta| > 2.0$) only pixel and SCT information is used.

The muon momentum is calculated by statistically combining the information from the ID and the MS, applying a parameterised correction for the energy loss in the calorimeter. Such muons are referred to as \emph{combined muons}. In some cases it is possible to match an ID track to a signal in the MS, but not possible to perform the combination because the MS track segment contains too few hits. In such cases, the ID track is used as an identified muon candidate. 
Muons that cross only the first layers of MS chambers, either due to
low transverse momentum or because they fall in an area of reduced MS
acceptance, can be identified in this less stringent
category. The inclusion of these \emph{segment-tagged muons} provides useful additional kinematic acceptance at low \pt\ for the reconstruction of particles with low invariant mass, such as the \Jpsi.

Muons originating from the $Z$ boson are required to be combined muons and have $p_\mathrm{T} > 15~\mathrm{GeV}$ and $|\eta|<2.5$. For the $J/\psi$ muons, one must be combined and the other can either be combined or segment-tagged. At least one of these two muons must have $p_\mathrm{T} > 4~\mathrm{GeV}$. Muons with $|\eta|>1.3$ are required to have $p_\mathrm{T} > 2.5~\mathrm{GeV}$ and muons with $|\eta|<1.3$ must have $p_\mathrm{T} > 3.5~\mathrm{GeV}$.

Electrons are reconstructed~\cite{Aad:2014nim} from energy deposits in the
electromagnetic calorimeter that are matched to a track in the inner detector. 
Candidate electron tracks are fitted using a dedicated tracking algorithm
to account for bremsstrahlung energy losses, and the track pattern recognition and global $\chi^2$ fit take
into account the electron track hypothesis as an alternative to the default pion hypothesis.  
Both electrons coming from the $Z$ boson decay need to have $p_\mathrm{T}>15\,\mathrm{GeV}$, $|\eta|< 2.47$ and satisfy the {\it loose identification} criteria described in Ref.~\cite{Aad:2014nim}.

In order to reject non-prompt leptons from the decay of heavy quarks, electrons from conversions of bremsstrahlung photons and fake electrons from misidentified jets, the leptons that form the $Z$ boson candidate must satisfy isolation requirements based on tracking information. The scalar sum of the transverse momenta 
of inner detector tracks inside an $\eta$--$\phi$ cone of size $\Delta R =0.2$ around the lepton, excluding the track associated with the lepton itself, is required to be no more than $15\%$ of the lepton $p_\mathrm{T}$.

At least one of the $Z$ boson candidate's leptons must have been responsible for firing the trigger. This criterion is assessed by requiring one of the reconstructed muons (electrons) from the boson to be less than $\Delta R < 0.1(0.15)$ from a relevant muon (electron) trigger object. The offline reconstructed $p_\mathrm{T}$ of the candidate matching the trigger must satisfy $p_\mathrm{T} > 25\,\mathrm{GeV}$. In addition, triggered muons must satisfy $|\eta|<2.4$ and electrons must satisfy the {\it medium identification} criteria, as described in Ref.~\cite{Aad:2014nim}.

\subsection{$Z\,+\,J/\psi$ candidate selection}

\begin{table*}[htpb]
\begin{center}
\caption{\label{tab:phasespace}Phase-space definition of the measured fiducial production cross-section following the geometrical acceptance of the ATLAS detector.}
\begin{tabular}{rl}
\hline\hline
\multicolumn{2}{c}{$Z$ boson selection}\\
\hline
   \\ [-2.0ex]
\pt(trigger lepton)$>25$~\GeV, & \pt(sub-leading lepton)$>15$~\GeV\\
\multicolumn{2}{c}{$|\eta(\mathrm{lepton~from}~Z)|<2.5$}\\
\multicolumn{2}{c}{$|m^Z-91.1876\,\GeV|<10$~\GeV}\\
   \\ [-2.0ex]
\hline
\multicolumn{2}{c}{\Jpsi selection}\\
\hline
   \\ [-2.0ex]
\multicolumn{2}{c}{$2.6<m^{J/\psi}<3.6\,\mathrm{GeV}$}\\
$8.5<\pt^{\Jpsi}<100\,\GeV$, & $|y_{\Jpsi}|<2.1$\\
\pt(leading muon)$>4.0$~\GeV, & $|\eta(\mathrm{leading~muon})|<2.5$\\
\ldelim\{{2}{5mm}either \pt(sub-leading muon)$>2.5$~\GeV, & $1.3\leq |\eta(\textrm{sub-leading~muon})|<2.5$ \rdelim\}{2}{2mm}[] \\
or \pt(sub-leading muon)$>3.5$~\GeV, & $|\eta(\textrm{sub-leading~muon})|<1.3$\\
   \\ [-2.0ex]
\hline\hline
\end{tabular}
\end{center}
\end{table*}

Same-flavour, opposite-sign lepton pairs are combined to reconstruct the $Z(\to \ell^+\ell^-)$ and $J/\psi(\to\mu^+\mu^-)$ candidates. Candidate \zjpsi\ events are retained if the \Jpsi\ invariant mass
falls in the range $2.6$--$3.6\,\mathrm{GeV}$ and the $Z$ boson candidate has an invariant mass within $10\,\mathrm{GeV}$ of the $Z$ mass world-average value ($m^Z_{\mathrm{PDG}}$)~\cite{Beringer:1900zz}.  
In addition, the
\Jpsi\ candidate is required to satisfy $p_\mathrm{T}>8.5\,\mathrm{GeV}$ and $|y(\Jpsi)|<2.1$. 
The measurements in the di-electron and di-muon decay channels of the $Z$ boson are performed in slightly different phase spaces and combined into a common phase-space 
for measurement of the fiducial production cross-sections as summarised in Table~\ref{tab:phasespace}.
The inclusive phase-space definition is identical except for the omission of requirements on the leptons from the \Jpsi decay.

The $Z$ boson and $J/\psi$ lepton pairs are used to build two dilepton vertices. In the case of the $J/\psi$ candidate the ID tracks alone are used in this vertex fit, whereas for the $Z\to\mu^+\mu^-$ the combined tracks (which are built from hits in both the ID and the MS) are used. For $Z\to e^+e^-$ decays, ID tracks corrected by a dedicated tracking algorithm are used, as described above. To reduce contamination from pileup, where a $Z$ boson and a $J/\psi$ are produced from two separate proton--proton collisions in the same proton--proton bunch crossing, the candidate vertices must not be separated in the $z$-direction by more than $10\,\mathrm{mm}$.

\begin{figure*}[tbp]
  \begin{center}
        \subfigure[] {
		\includegraphics[width=0.99\columnwidth]{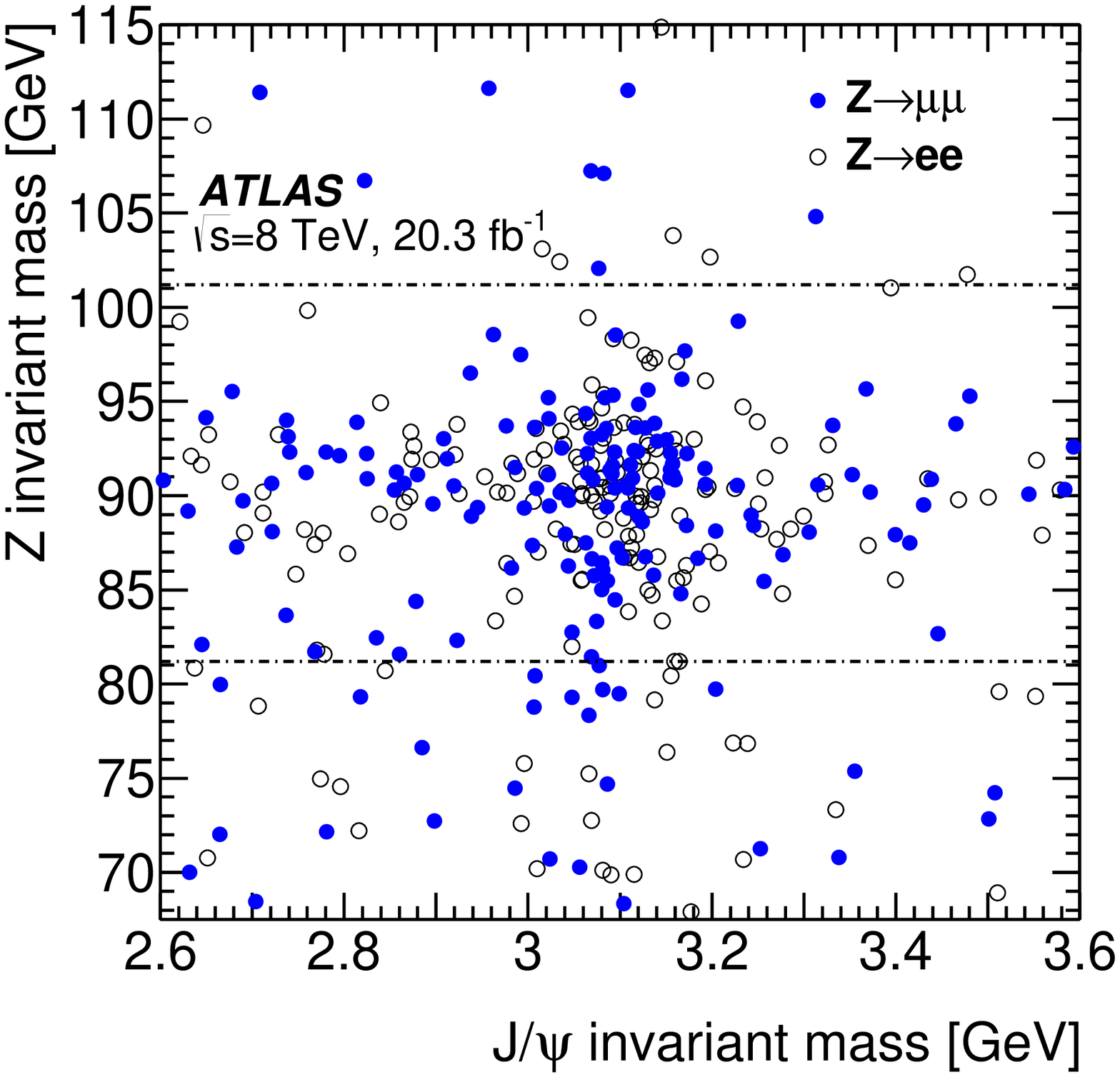}
	          \label{fig:2dhistsmass}
        }
        \subfigure[] {
		\includegraphics[width=0.99\columnwidth]{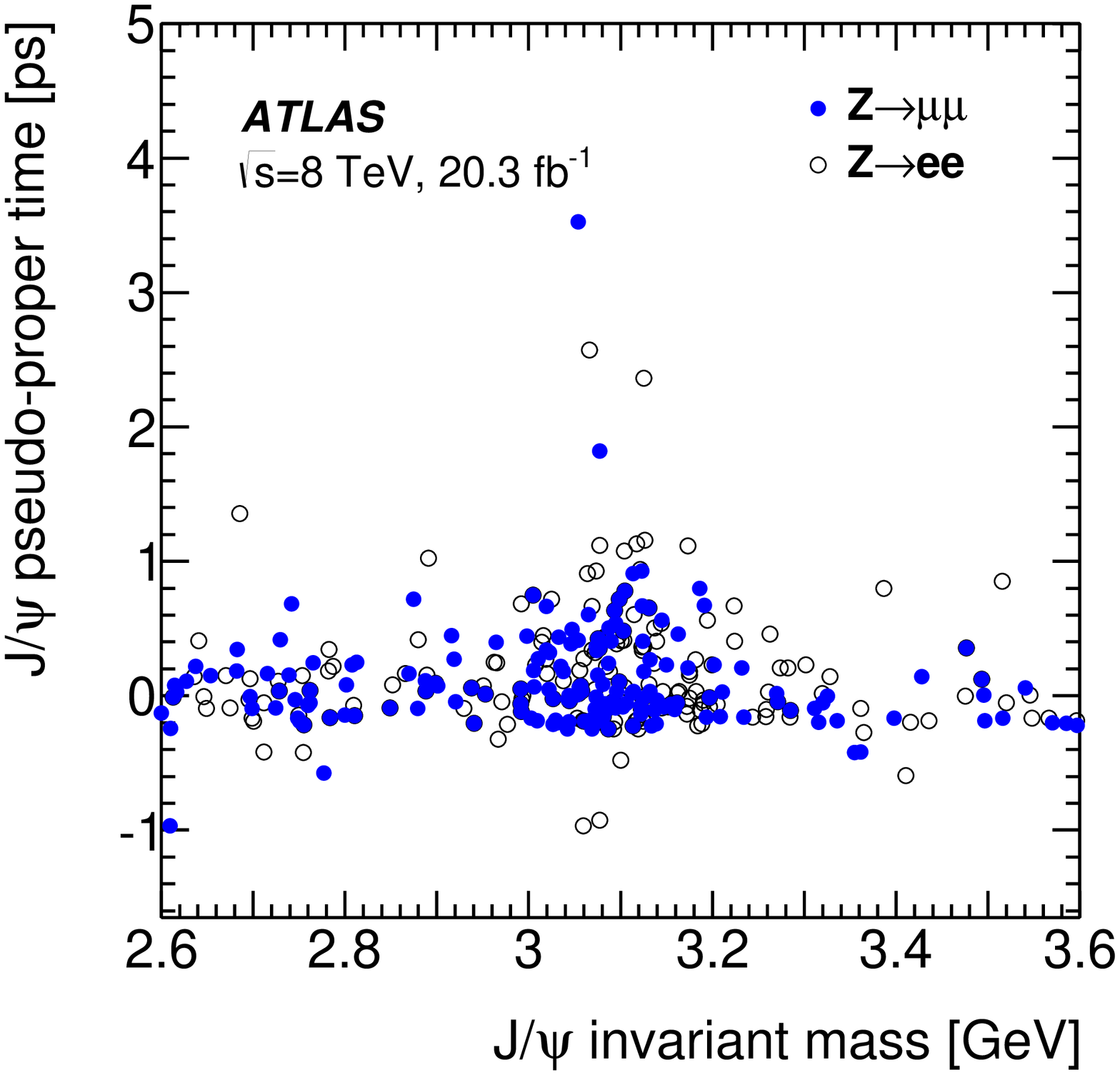}
	          \label{fig:2dhiststime}
        }
    \caption{Selected $Z\,+\,J/\psi$ candidates in (a) $Z$ boson mass versus $J/\psi$ boson mass, with $\ell=e,\mu$ and (b) $\Jpsi$ pseudo-proper time versus \Jpsi invariant mass, discussed in Section~\ref{sec:promptnonpromptJpsiseparation}. 
      $Z$ boson candidates decaying to muons are shown with full circles and to electrons with empty circles. The horizontal dotted lines indicate the signal region considered in the analysis.
      \label{fig:2dhists}
   }
  \end{center}
\end{figure*}

Figure~\ref{fig:2dhistsmass} shows a scatter plot of the masses of candidates satisfying these selections. 
In total, $290$ candidate events are selected, of which $139$ are observed with $Z\to\mu^+\mu^-$ decays and $151$ with $Z\to e^+e^-$ decays.

\subsection{Inclusive $Z$ candidate selection} \label{sec:eventselectionInclusiveZ}

An inclusive $Z$ sample is formed by selecting all events that satisfy
the $Z$ part of the $Z\,+\,J/\psi$ selection, including the trigger
requirements. This sample is used in the measurement of the ratio of
the $Z\,+\,J/\psi$ to $Z$ cross-sections, and in the estimates of double
parton scattering and the pileup background in the associated-production
sample.

An estimate of the background in the inclusive $Z$ sample is obtained
using a mixture of Monte Carlo (MC) models and data-driven techniques.
The NLO generator {\sc Powheg (r1556)}~\cite{Nason:2004rx,Frixione:2007nw,Frixione:2007vw}, 
interfaced to {\sc Pythia (8.160)}~\cite{pythia8}, is
used to model the signal, as well as Drell--Yan contributions away from
the $Z$ peak and $Z\to\tau\tau$ or $W\to\lnu$ backgrounds. These samples use the CT10 PDF set~\cite{Lai:2010vv} and the ATLAS AU2 tune~\cite{ATLAS:2011zja}. The LO multi-leg generator {\sc Sherpa v1.4.1}~\cite{Gleisberg:2008ta} is used as an
alternative signal model.  Top quark processes involving \ttbar\ or
single top production are modelled with the NLO generator {\sc MC@NLO (4.03)}~\cite{Frixione:2002ik,Frixione:2003ei}, interfaced to {\sc Herwig (6.52)}~\cite{Corcella:2000bw} for parton showering and {\sc Jimmy (4.31)}~\cite{Butterworth:1996zw} for the underlying-event modelling with the ATLAS AUET2 tune~\cite{ATLAS:2011gmi} and the CT10 PDFs. The single-top $Wt$
process is modelled with the {\sc AcerMC (3.8)}~\cite{Kersevan:2004yg} generator, using the CTEQ6L1 PDF set~\cite{Pumplin:2002vw} and interfaced to {\sc Pythia (6.42)}~\cite{Sjostrand:2006za}.  Diboson ($WZ$, $WW$ and $ZZ$) production is modelled using the {\sc Herwig (6.52)}  and Jimmy generators with the ATLAS AUET2 tune and the CTEQ6L1 PDF set. The detector response is modelled using the ATLAS simulation infrastructure~\cite{Aad:2010ah} based on the {\sc Geant4} toolkit~\cite{Agostinelli:2002hh}.  
Background contributions arising from
multi-jet events and from misidentified leptons are obtained directly from the data. This is achieved by inverting
the isolation requirements on the leptons, providing a multi-jet
background template, which can be used for comparison with the
$Z\,+\,J/\psi$ sample. 
The total background in the $m^Z_\mathrm{PDG}\pm10\,\mathrm{GeV}$ window is estimated to be $0.4\pm 0.4\%$ (including systematic uncertainties), giving a sample of $16.15$ million $Z$ boson candidates after background subtraction, of which $8.20$ million are observed with $Z\to\mu^+\mu^-$ and $7.95$ million with $Z\to e^+e^-$. The di-muon to di-electron ratios of the associated-production $Z\,+\,J/\psi$ sample and the inclusive $Z$ sample are compared and found to be consistent within statistical uncertainties ($0.92\pm 0.11$ and $1.03\pm 0.01$, respectively).

\section{Signal and background extraction} \label{sec:signalextraction}

The selected \zjpsi candidates arise from a variety of signal and background sources. 
In addition to associated $Z$ boson and $J/\psi$ production from SPS and DPS, $Z$ boson 
and $J/\psi$ candidates can be produced from pileup. 
Genuine $J/\psi$ may also be paired with fake $Z$ boson candidates in the same proton--proton collision, or vice-versa. 
Associated-production candidates may also occur from the production of a $Z$ boson in association with $b$-quarks, where one of the $b$-quarks hadronises into a $b$-hadron that subsequently decays into a $J/\psi$. 
This section discusses the means by which the contributions from the prompt and non-prompt signal components are distinguished and separated from the prompt and non-prompt background sources.

\subsection{Separation of prompt and non-prompt $J/\psi$} \label{sec:promptnonpromptJpsiseparation}

The $J/\psi\to\mu^+\mu^-$ candidates originate from prompt and non-prompt production sources, backgrounds with real and fake muon combinations, and real muon pairs producing an invariant mass in the continuum under the $J/\psi$ peak. 
These various components can be separated into categories using the pseudo-proper 
time distribution of the $J/\psi$ candidates in combination with the \Jpsi invariant mass distribution, where the pseudo-proper time, $\tau$, is defined by:

\begin{equation}
\tau\coloneqq\frac{L_{xy}m^{J/\psi}}{p_\mathrm{T}^{J/\psi}}
\end{equation}
with $L_{xy}$ defined as
$L_{xy}=\vec{L}\cdot\vec{p_\mathrm{T}}^{J/\psi}/p^{J/\psi}_\mathrm{T}$,
$\vec{L}$ the vector from the primary vertex to the $J/\psi$ decay
vertex, $m^{J/\psi}$ the world-average mass of the
$J/\psi$ meson~\cite{Beringer:1900zz}, $\vec{p_\mathrm{T}}^{J/\psi}$ the transverse momentum of the $J/\psi$ and $p_\mathrm{T}^{J/\psi}=|\vec{p_\mathrm{T}}^{J/\psi}|$ its magnitude. The invariant mass and pseudo-proper time of the selected $J/\psi$ candidates produced in association with a $Z$ boson are shown in Fig.~\ref{fig:2dhiststime}.

Promptly produced $J/\psi$ mesons, which are created directly in the hard interaction or feed-down from prompt excited charmonium states produced by the colliding protons, have small pseudo-proper times (distributed around zero due to detector resolution). 
Background from opposite-sign muon pairs with invariant mass close to the $J/\psi$ mass and short reconstructed pseudo-proper times can mimic prompt $J/\psi$ mesons and forms the prompt background. 
The second component of the background arises from non-prompt muon pairs, with a vertex displacement that is related to $b$-hadron decays. 
Similarly, the signal from non-prompt $J/\psi$ production exhibits a
longer pseudo-proper time distribution reflecting the lifetime of $b$-hadrons, although the distributions of non-prompt signal and
background are not necessarily equal. In total, four terms are used for signal and background to fit the pseudo-proper time distribution simultaneously with the invariant mass distribution of the muon pair. The mass regions either side of the $J/\psi$ mass peak are used to constrain the background components.

The pseudo-proper time of the signal prompt component is modelled by a double Gaussian distribution. For the background prompt component, a double-sided exponential convolved with the prompt signal function, accounting for resolution effects, is used. 
The non-prompt signal component is modelled with a single-sided exponential convolved with the prompt signal function and for the non-prompt background component the sum of a single-sided and a double-sided 
exponential convolved with the signal function is used. The dimuon invariant mass is modelled with a double Gaussian distribution both for the prompt and non-prompt signal components and 
exponential functions for the backgrounds (again, prompt and non-prompt). The fit is performed in two separate rapidity regions, 
the barrel ($|y_{J/\psi}|<1.0$) and the endcap ($1.0<|y_{J/\psi}|<2.1$). The mass resolution is different between the two regions, due to increased multiple scattering and the decrease of the magnetic field integral at high rapidity. 

\begin{figure*}[htbp]
  \begin{center}
        \subfigure[] {
		\includegraphics[width=0.99\columnwidth]{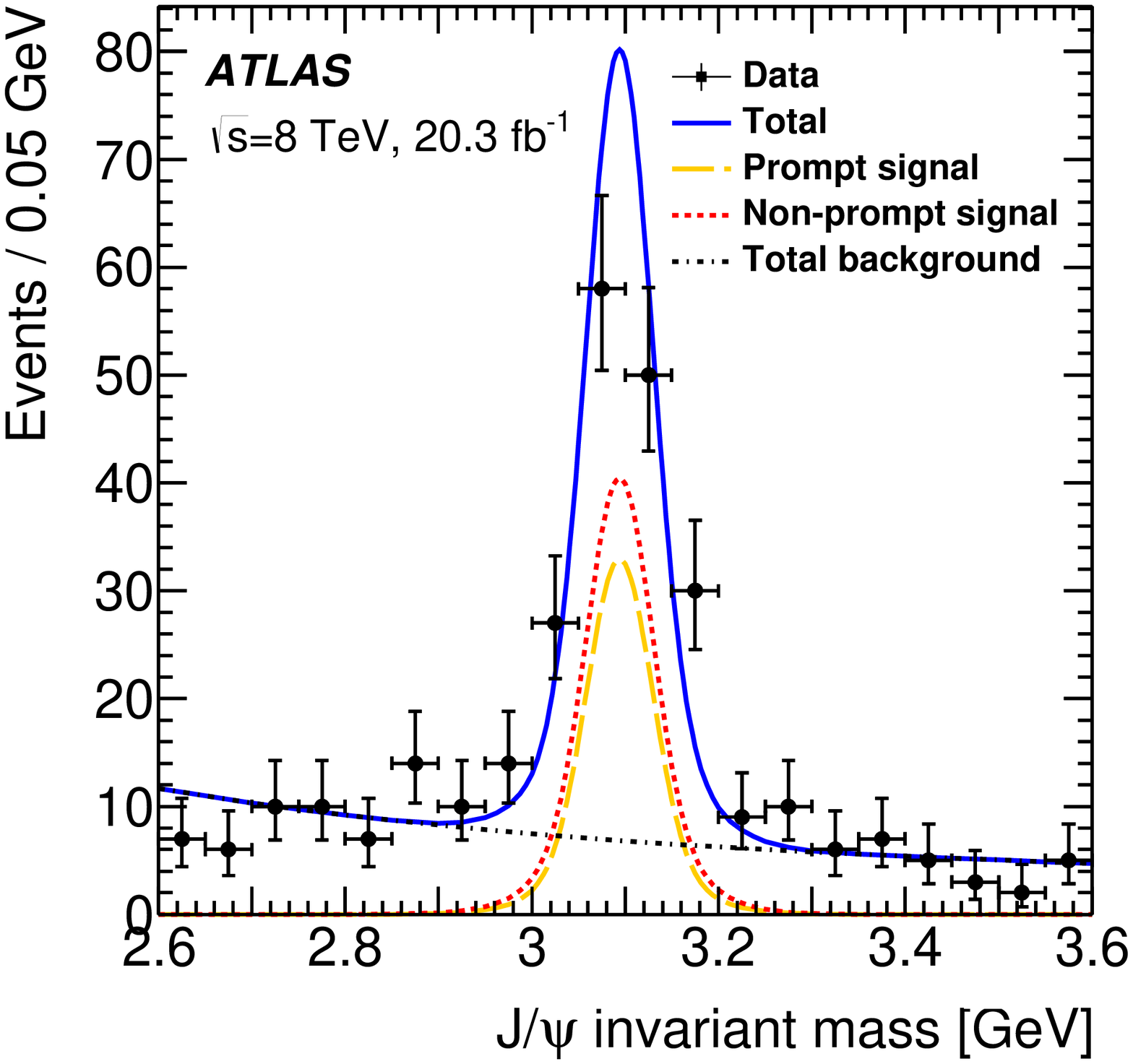}
	          \label{fig:cross_section_jpsi_mass}
        }
        \subfigure[] {
		\includegraphics[width=0.99\columnwidth]{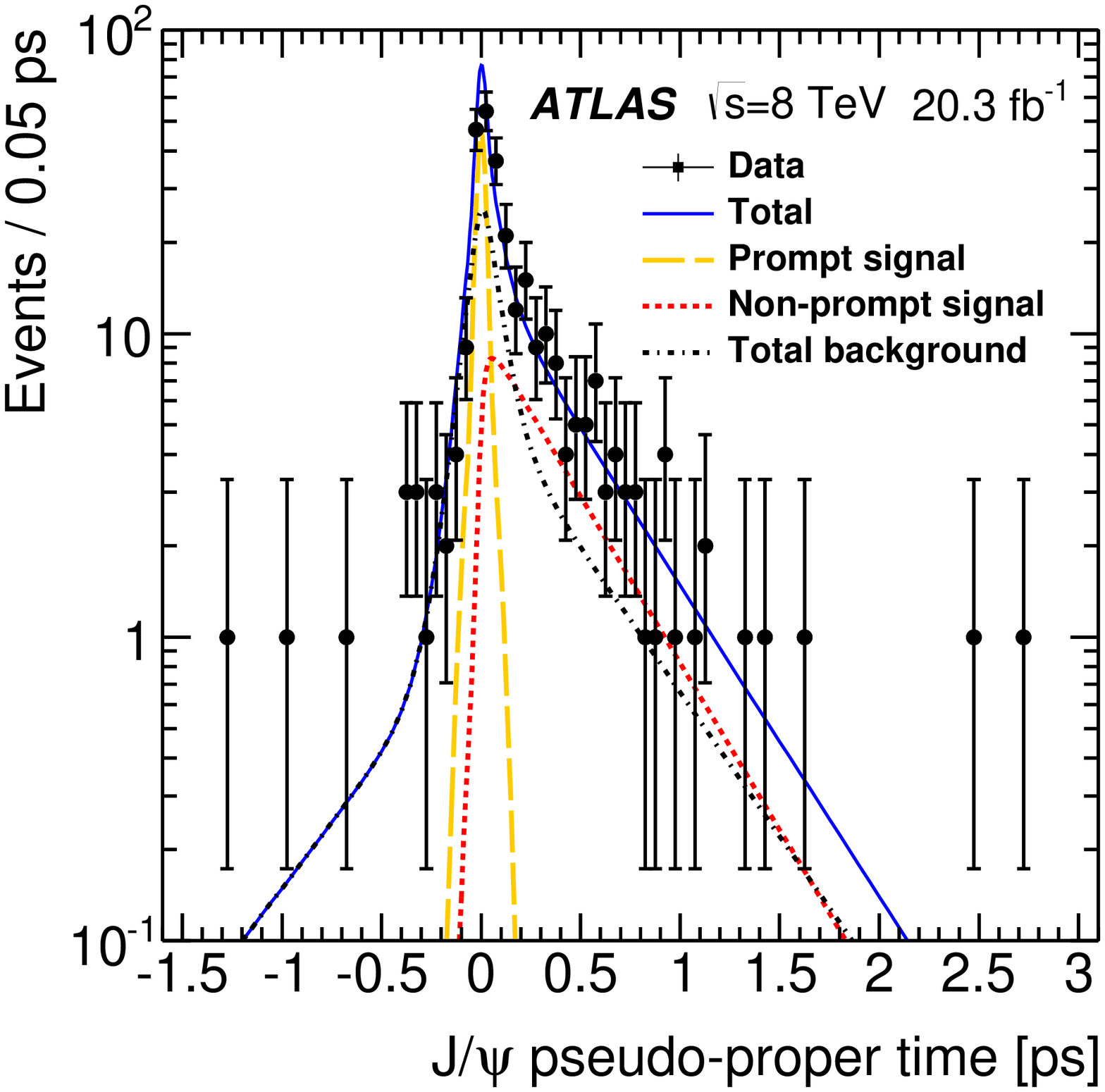}
	          \label{fig:cross_section_jpsi_time}
        }
    \caption{Projections of the unbinned mass and pseudo-proper time maximum-likelihood fit in (a) invariant mass and (b) pseudo-proper time of the associated-production sample. The fit is used to extract the prompt and non-prompt signal fractions and is performed in two rapidity regions: $|y_{J/\psi}|<1.0$ and $1.0<|y_{J/\psi}|<2.1$. The results are combined, presenting the mass and pseudo-proper time of all candidates inside the analysis phase-space.
      \label{fig:cross_section_jpsi}
    }
  \end{center}
\end{figure*}

In order to improve the stability of the fit process, the pseudo-proper time and invariant mass of the associated-production $J/\psi$ candidates are fitted simultaneously with a sample of $100\,\mathrm{k}$ inclusive $J/\psi$ candidates, selected with the same requirements on the $J/\psi$ and its daughter muons as applied to the $Z\,+\,J/\psi$ signal sample (see Table~\ref{tab:phasespace}). 
The parameters that determine the shape of the pseudo-proper time and invariant mass distributions are linked between the two samples in this fit, leaving only the normalisations free between the two samples. 
Figure~\ref{fig:cross_section_jpsi} shows the mass and pseudo-proper time distributions of the $J/\psi$ candidates, produced in association with a $Z$ boson, with the signal and background fits. Applying the fit model to the sample of $Z$ bosons produced in association with a $J/\psi$ candidate results in $56\pm 10$ promptly produced $J/\psi$ mesons and $95\pm 12$ non-prompt.

After the fit is performed in the $J/\psi$ mass and pseudo-proper time,
the sPlot tool~\cite{splot} is used to assign a weight to each event for each of the components included in the fit model (prompt signal, non-prompt signal, prompt background and non-prompt
background). This technique allows the determination of distributions of observables associated with a specific contribution, e.g. prompt $J/\psi$, while removing the contamination from the other components. As the sPlot technique relies on the assumption that the control variable is uncorrelated with the discriminating variables, the correlations between the $J/\psi$ mass and pseudo-proper time, on one side, and the variables that the weights will be applied to, on the other, were checked, and found to be negligible. The invariant mass distribution of $Z$ boson candidates, after application of the sPlot weights, is shown in
Figs.~\ref{fig:z_splots_prompt_main}
and~\ref{fig:cross_section_z_nonprompt} for prompt $J/\psi$ and
non-prompt $J/\psi$ events, respectively.

\subsection{Properties of the $Z$ boson candidates}
\begin{figure*}[htb]
	\begin{center}
		\subfigure[$Z\to e^+e^-$ (left) and $Z\to\mu^+\mu^-$ (right) associated with prompt $J/\psi$.]{\includegraphics[width=0.92\textwidth]{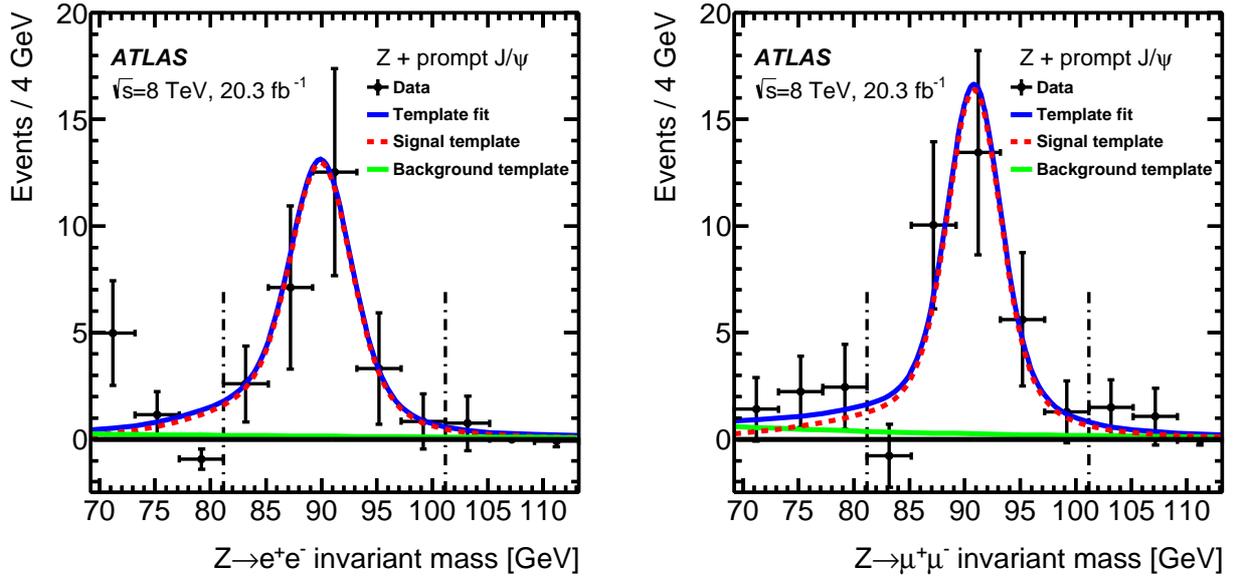}
		\label{fig:z_splots_prompt_main}}
		\subfigure[$Z\to e^+e^-$ (left) and $Z\to\mu^+\mu^-$ (right) associated with non-prompt $J/\psi$.]{\includegraphics[width=0.92\textwidth]{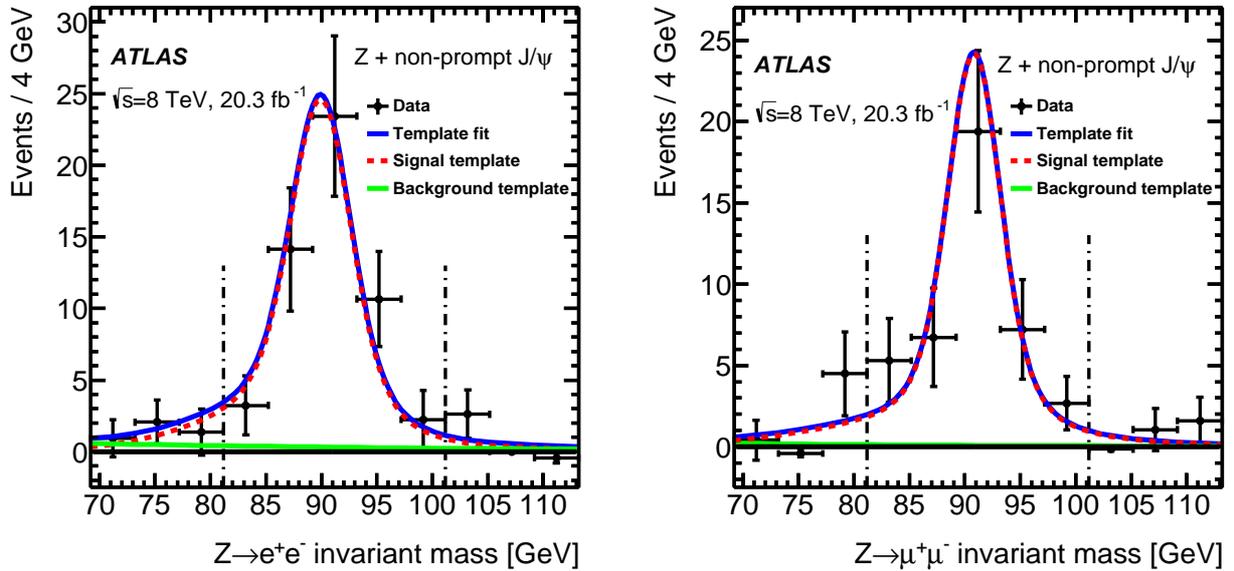}
		\label{fig:cross_section_z_nonprompt}}
		\caption{$Z\to e^+e^-$ (left) and $Z\to\mu^+\mu^-$ (right) candidate invariant mass distributions after the application of the sPlot weights coming from the (a) prompt and (b) non-prompt $J/\psi$ component of the fit. Projections of the unbinned maximum likelihood template fit, for the signal and background components derived from MC simulation and data respectively, are overlaid on the sPlot-weighted distributions. The vertical dot-dashed lines indicate the signal region considered in the analysis.}
 		\label{fig:cross_section_z_all}
	\end{center}
\end{figure*}

Signal and multi-jet background templates for the dilepton mass were extracted separately for $Z\to\ee$ and
$Z\to \mumu$ from the {\sc Powheg} MC generator described in
Sect.~\ref{sec:eventselectionInclusiveZ} and the data.  The signal templates are parameterised with a
Gaussian distribution convolved with a Breit--Wigner function, with an
additional Gaussian, with smaller mean value compared to the core Gaussian, to model the radiative tails. The multi-jet templates are modelled with an exponential function. The normalisations of the two templates are extracted from a fit to the sPlot-weighted $Z$ invariant mass distributions
(Fig.~\ref{fig:cross_section_z_all}). The numbers of background events estimated in the $Z$ signal region, defined as $m_\mathrm{PDG}^{Z}\pm 10\,\mathrm{GeV}$, are $0\pm 4\ (1\pm 4)$ and $1\pm 5\ (0\pm 5)$
for the $Z\to e^+e^-(\mu^+\mu^-)$ candidates associated with prompt and
non-prompt $J/\psi$, respectively, supporting the hypothesis that the sample is dominated by genuine $Z\,+\,J/\psi$ events. The background estimation procedure was verified with toy MC simulation.

\subsection{Pileup background} \label{sec:pileupmain}
During the 2012 data-taking period the average number of $pp$ interactions per
bunch crossing at ATLAS was $20.7$. While the
most likely scenario is that all but one of these inelastic collisions
are low-$p_\mathrm{T}$ background events, there is a certain
probability that two or more of these produce a hard scatter. Of these
cases, some produce a $Z$ from one scatter, and a $J/\psi$ from
another. To exclude as many as possible of these background events,
the two dilepton vertices are required to be separated along the $z$-axis by less than $10\,\mathrm{mm}$. The remaining contamination can be estimated using four ingredients: the spread of the beam spot
in $z$ for the data-taking period of relevance; the $J/\psi$
production cross-sections (prompt or non-prompt) from $pp$ collisions
at $8\,\mathrm{TeV}$; the number of $Z$ candidates; and the mean
number of inelastic interactions per proton--proton bunch crossing,
$\langle\mu\rangle$.  This latter quantity is calculated from the
instantaneous luminosity, $\mathcal{L}$, as $\langle\mu\rangle=\mathcal{L}\sigma_\mathrm{inel}/n_{\rm b} f_{\rm r}$, where $\sigma_\mathrm{inel}$ is the $pp$ inelastic cross-section (equal to 73~mb~\cite{Aad:2011dr}), $n_{\rm b}$ is the number of colliding bunches and $f_{\rm r}$ is the LHC revolution frequency.

To estimate the mean number of pileup collisions occurring within $10\,\mathrm{mm}$ of a given $Z$ vertex, an MC procedure is used. A number of pileup vertices are sampled from the luminosity-weighted distribution of $\langle\mu\rangle$.  These vertices are distributed according to a Gaussian function with width $48\pm 3\,\mathrm{mm}$, equal to the measured width of the proton beam spread in the $z$-coordinate. The number of additional vertices which lie within $10\,\mathrm{mm}$ of a randomly selected vertex, is determined to be $N_{\mathrm{extra}} = 2.3\pm 0.2$.

As it has been verified that the $J/\psi$ reconstruction efficiency is independent of the number of interactions per bunch crossing, the probability for a \jpsi\ to be produced at a given pileup vertex is

\begin{equation}
P^{ij}_{J/\psi}=\sigma^{ij}_{J/\psi}/\sigma_{\mathrm{inel}}
\label{eq:sigmapileup}
\end{equation}

\noindent where $\sigma^{ij}_{J/\psi}$ is the cross-section for \Jpsi production in the appropriate \pt\ ($i$) and rapidity ($j$) bin.  Although $\sigma^{ij}_{J/\psi}$ has not been measured in the fiducial region used in this measurement at centre-of-mass energies of $\sqrt{s}=8\,\mathrm{TeV}$, it can be estimated using an existing non-prompt $J/\psi$ fraction measurement at $\sqrt{s}=7\,\mathrm{TeV}$~\cite{JpsiATLAS7TeV} and the fixed-order next-to-leading-logarithm~\cite{FONLL,Cacciari:2012ny} (FONLL) prediction for the non-prompt $J/\psi$ cross-section at $\sqrt{s}=8\,\mathrm{TeV}$. This extrapolation to $8\,\mathrm{TeV}$ is based on the observation~\cite{JpsiATLAS7TeV} that the variation in the ratio of non-prompt to prompt $J/\psi$ production with \pt\ appears to be independent of the collision energy, and also on the excellent agreement between the ATLAS measurement and the FONLL predictions of the non-prompt cross-section. 

The number of pileup candidates can be evaluated using the number of $Z$ candidates in the fiducial region, $N_Z$, according to $N^{ij}_{\mathrm{pileup}} = N_{\mathrm{extra}}N_ZP^{ij}_{J/\psi}$, giving a total of $\sum_{i,j}N^{ij}_{\mathrm{pileup}}=5.2^{+1.8}_{-1.3}$ and $2.7^{+0.9}_{-0.6}$ events in the prompt and non-prompt samples, respectively. The uncertainty on the final result includes contributions from the estimated $J/\psi$ cross-section at $\sqrt{s}=8\,\mathrm{TeV}$, the number of inclusive $Z$ events and the number of extra vertices. The dependence of $\langle\mu\rangle$ and $P_{J/\psi}$ on $\sigma_{\mathrm{inel}}$ cancels in the determination of $N_{\mathrm{pileup}}$. 

\subsection{Double parton scattering}

The DPS contribution to the $Z\,+\,J/\psi$ sample is counted as part of the signal. The effective cross-section for double parton interactions $\sigma_\mathrm{eff}$ measured  by ATLAS in $W\,+\,2$-jet events~\cite{ATLASsigmaeff}, and the $pp\to J/\psi$ prompt and non-prompt cross-sections, are used to estimate the number of signal candidates from this source. Based on the assumptions that $\sigma_\mathrm{eff}$ is process-independent, and that the two hard scatters are uncorrelated, for a collision where a $Z$ boson is produced, the probability that a $J/\psi$ is produced in addition due to a second hard process is 

\begin{equation}
P^{ij}_{J/\psi|Z}=\sigma^{ij}_{J/\psi}/\sigma_\mathrm{eff}
\label{eq:sigmaeff}
\end{equation}

\noindent where $\sigma_{\mathrm{eff}}$ is taken to be $\sigma_{\mathrm{eff}}=15\pm 3\,(\mathrm{stat.}) ^{+5}_{-3}\,\mathrm{(sys.)}\,\mathrm{mb}$ according to the ATLAS measurement. 
The estimated numbers of DPS events in the associated-production $Z\,+\,J/\psi$ sample are $11.1^{+5.7}_{-5.0}$ for the prompt component and $5.8^{+2.8}_{-2.6}$ for the non-prompt component. Uncertainties from the $J/\psi$ cross-section at $\sqrt{s}=8\,\mathrm{TeV}$, the number of inclusive $Z$ events and the DPS effective cross-section contribute to the total uncertainty.

\begin{figure*}[thbp]
  \begin{center}
\includegraphics[width=0.99\columnwidth]{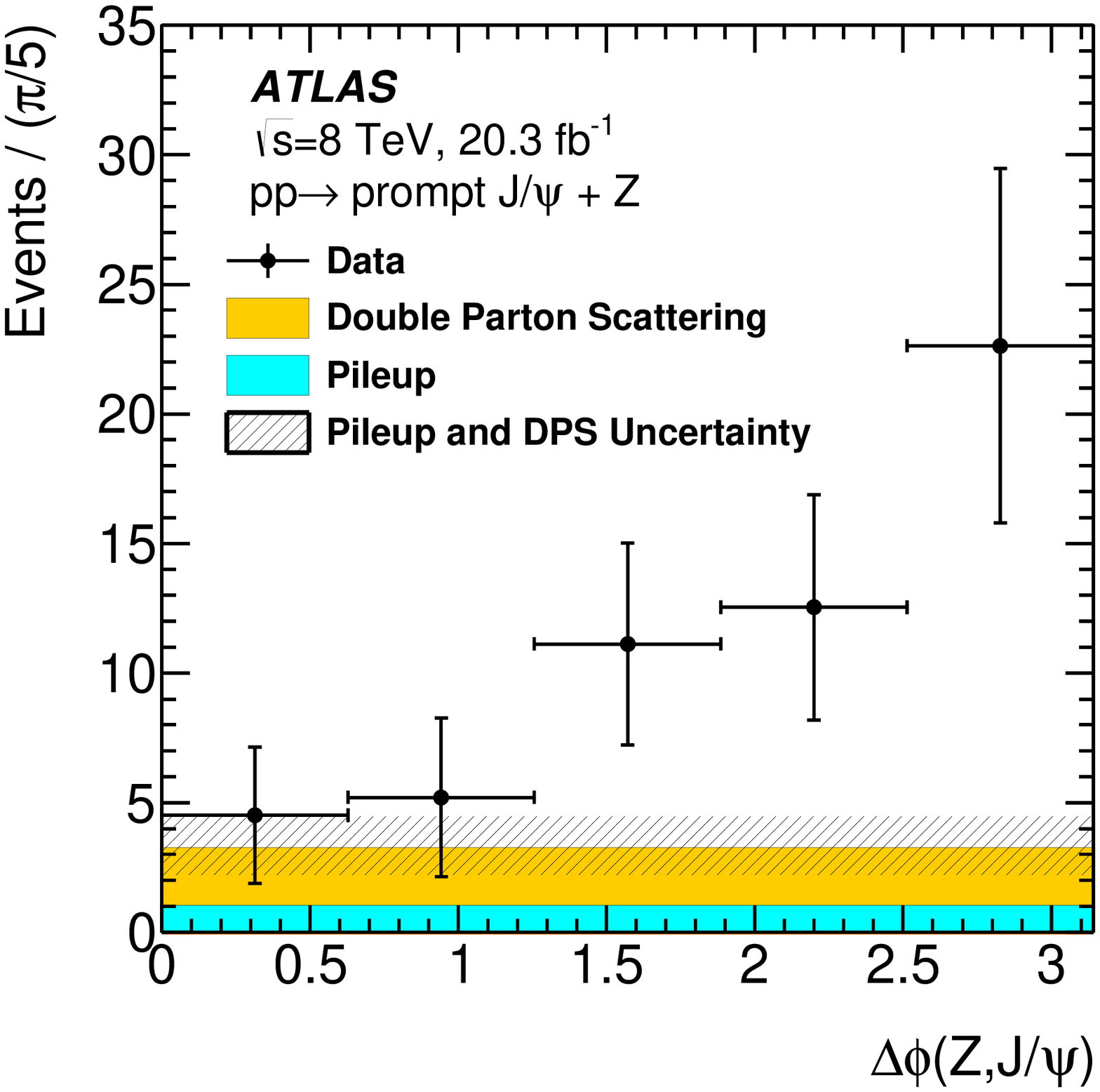}\label{fig:sPlot_Delta_phi_all_prompt}
\includegraphics[width=0.99\columnwidth]{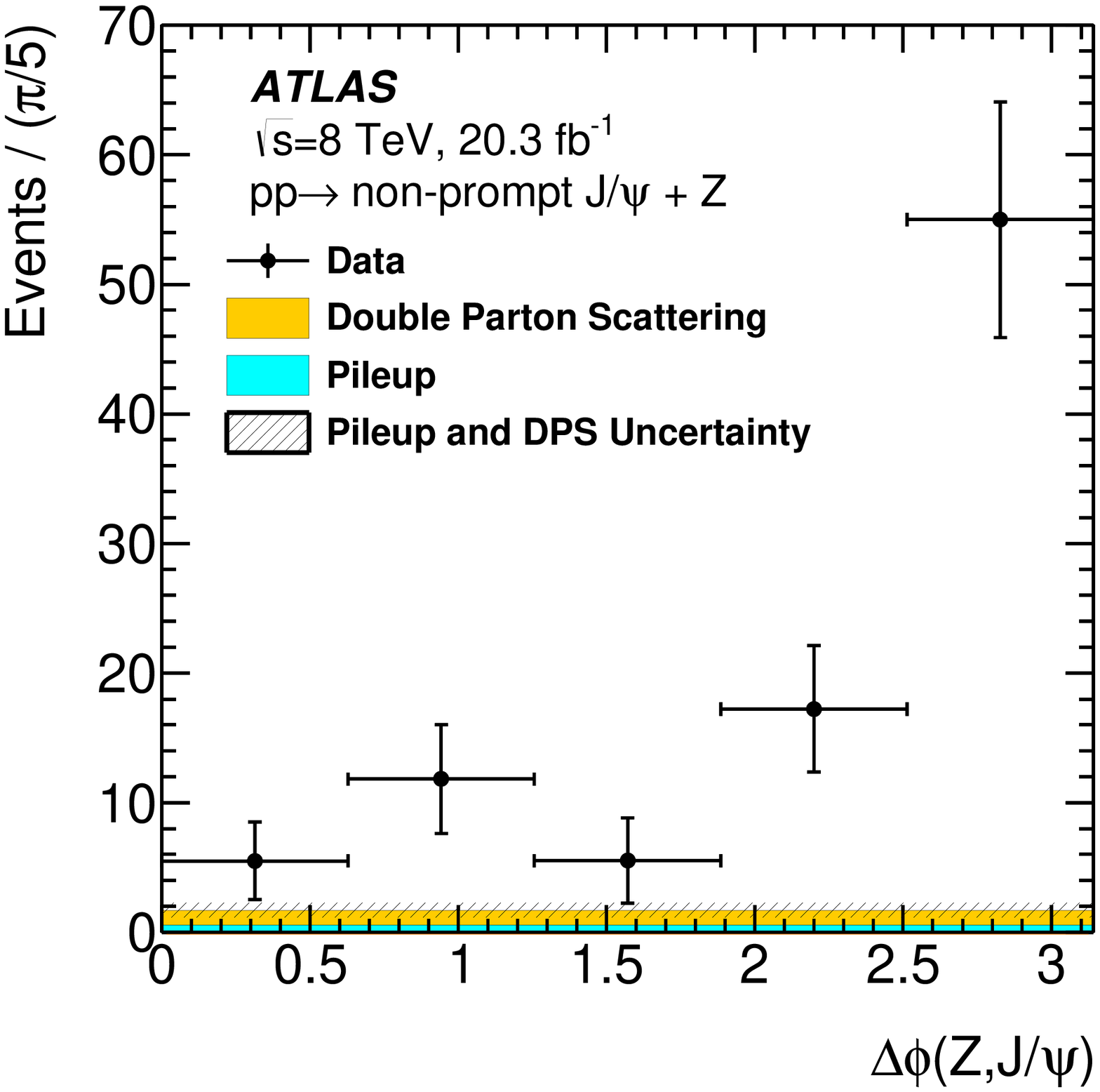}\label{fig:sPlot_Delta_phi_all_nonprompt}
    \caption{Azimuthal angle between the $Z$ boson and the $J/\psi$ meson after the application of the sPlot weights to separate the 
prompt (left) and 
non-prompt (right) yield from background contributions. 
The estimated DPS (yellow band) and pileup (cyan band) contributions to the observed data are overlaid. The hashed region show the DPS and pileup uncertanties added in quadrature. 
    \label{fig:deltaphi}
    }
  \end{center}
\end{figure*}

Figure~\ref{fig:deltaphi} shows the azimuthal angle between the $Z$ boson and the $J/\psi$ momentum vectors, $\Delta\phi$, after the application of sPlot weights to separate the prompt and non-prompt $J/\psi$ signal components from each other and from background sources. 
The estimated contributions of double parton scattering and pileup to the
observed signal yields for prompt and non-prompt production
are also overlaid. DPS events  are expected
to be distributed uniformly in $\Delta\phi$ because the $Z$ and the
$J/\psi$ are produced by two independent processes. On the contrary,
SPS events are expected to display a back-to-back correlation of the
$Z$ and the $J/\psi$ ($\Delta\phi=\pi$) since the two particles come
from a single interaction of two partons. 
This back-to-back behaviour is smeared by the presence of
additional gluons in the final state, radiation from the leptons,
detector effects and by the intrinsic properties of the protons; the measured data are consistent with a combination of a smeared $\Delta\phi=\pi$ peak from SPS and a flat DPS contribution with $\sigma_\textrm{eff}$ taken from the ATLAS $W\,+\,2$-jet measurement.

\subsection{Detector effects and acceptance corrections}
The efficiency for reconstructing muons in the ATLAS detector is very high~\cite{muons} and depends on the kinematics of the muon. In order to correct the measurements for detector effects, a per-event weight is applied, based on the pseudorapidity and transverse momentum of both muons coming from the $J/\psi$ decay. These weights are extracted using large inclusive $J/\psi\to\mu^+\mu^-$ and $Z\to\mu^+\mu^-$ data samples and have been validated with MC simulation~\cite{muons}. Small inefficiencies resulting from the requirement on separation of $Z$ and $J/\psi$ vertices are corrected using MC simulations.

It was verified using MC simulation that detector resolution effects causing reconstructed $Z$ boson candidates to migrate in and out of the phase space defined in Table~\ref{tab:phasespace} do not produce visible effects on the measured relative production rates.

In addition to corrections applied for reconstruction efficiency (approximately $90\%$ depending on the $p_\mathrm{T}$ of the $J/\psi$), the detector acceptance needs to be taken into account.
The spin-alignment profile of the $J/\psi$ meson produced in  association with a $Z$ boson might be different from the known profile of inclusive $J/\psi$ mesons~\cite{Chatrchyan:2013cla}.
The modified angular distributions of muons from the decay of alternatively-polarised $\Jpsi$ mesons can cause changes in acceptance in the fiducial region defined by the selection requirements (see Table~\ref{tab:phasespace}). For various extreme polarisation states of the $J/\psi$ \cite{Faccioli:2010kd}, the $J/\psi$ rate is corrected for muons that fall outside the detector acceptance in transverse momentum and pseudorapidity.

\section{Systematic uncertainties} \label{sec:systematics}

\begin{table*}[thbp]
\begin{center}
  \caption{Summary of experimental systematic uncertainties.}
  \begin{tabular}{ l  c  c c c}
    \hline
     \multirow{2}{*}{Source} & \multicolumn{2}{ c }{Prompt}& \multicolumn{2}{ c }{Non-prompt} \\
    & $|y_{J/\psi}|<1.0$  & $1.0<|y_{J/\psi}|<2.1$ & $|y_{J/\psi}|<1.0$  & $1.0<|y_{J/\psi}|<2.1$   \\
\hline
   \\ [-2.0ex]
    Fit procedure & $3\,\%$ & $3\,\%$& $4\%$ & $8\%$ \\
    $Z$ boson kinematics & $1\%$  & $1\%$& $1\%$  & $1\%$ \\
    $\mu_{J/\psi}$ efficiency & $1\%$ & $1\%$ & $1\%$ & $1\%$\\
    Vertex separation & $7\%$ & $16\%$ & $2\%$ & $15\%$\\
   \\ [-2.0ex]
\hline
  \end{tabular}
  \label{tab:summary_of_systematics}
\end{center}
\end{table*}

Systematic uncertainties coming from the fit are calculated by varying the probability density functions for the \Jpsi\ mass 
and pseudo-proper time distributions. In addition to the model described in Sect.~\ref{sec:signalextraction}, an alternative model was used, changing the parameterisation for the mass and lifetime resolution and the shapes of the background components. This model parameterised the mass with a Gaussian function for the $J/\psi$ signal and exponential (or polynomial) functions for the combinatorial background, 
and parameterised the pseudo-proper time with the sum of a Gaussian and a double-sided exponential function convolved with a Gaussian resolution function for
the prompt $J/\psi$ and prompt combinatorial background component, and an exponential function convolved with a Gaussian resolution function for the 
non-prompt $J/\psi$ and non-prompt combinatorial background. The shape-related parameters are linked between the $Z\,+\,J/\psi$ sample and the 
inclusive $J/\psi$ sample in the model used for the signal extraction. This assumption neglects the possible difference in kinematics between $J/\psi$ 
mesons that are produced inclusively and $J/\psi$ mesons produced in association with a $Z$ boson and needs to be taken into account. This effect is 
evaluated by removing the link between the parameters and repeating the fit, using the main fit model and the alternative considered for the systematic study. 
The systematic uncertainty associated with the fit procedure was determined with a toy MC simulation
technique. A large number of simulated data samples were generated for the two rapidity bins and then fitted with all the available fit procedures. 
The uncertainties were evaluated from the maximal variation in mean yield extracted from each of the three fit models, relative to the nominal model.
This uncertainty was found to be $3\%$ for prompt production and $4\%$--$8\%$ (depending on the rapidity of the \Jpsi\ candidate) for non-prompt production.

In the measurement of the cross-section ratios, it is assumed that the efficiency and acceptance
for the $Z$ boson are the same when the $Z$ is produced in association
with a $J/\psi$ as when it is produced inclusively. 
In the absence of reliable signal Monte Carlo samples for the SPS or DPS processes, 
systematic uncertainties that arise from this assumption are calculated using a data-driven approach.
The reconstruction and trigger efficiencies calculated for the associated-production data sample
and an inclusive $Z$ sample, re-weighted to match the observed $Z\,+\,J/\psi$ $p_\mathrm{T}$ spectrum, are compared. 
The non-cancellation of efficiencies and acceptance between inclusively-produced $Z$ bosons and those produced in association with a \Jpsi\ is found to be $(1\pm 1)\%$. 

The reconstruction efficiencies of the $J/\psi$ muons used for the correction and calculation of the inclusive cross-section are extracted from   $Z\to\mu^+\mu^-$ and $J/\psi\to\mu^+\mu^-$ decays using a tag-and-probe method~\cite{muons}. These efficiencies and their uncertainties depend on the muon pseudorapidity and $p_\mathrm{T}$ and are applied to the data in the form of two-dimensional maps. In order to calculate the systematic uncertainty, each bin of the efficiency map is allowed to vary within its uncertainty and the effect on the extracted yield is examined. The systematic uncertainty from the muon reconstruction efficiency is of the order of $1\%$.

In the selection requirements applied to the dataset, the $Z$ and $J/\psi$ vertices are required to be within $10\,\mathrm{mm}$ along the $z$-axis. This choice could cause a potential bias in the measurement of the prompt and the non-prompt yield since it affects the pseudo-proper time distribution of the $J/\psi$. This cut is loosened to $20\,\mathrm{mm}$ and the difference in the extracted yield, again assessed using data-driven pseudoexperiments, determined after the pileup subtraction and correction for the expected change in signal efficiency from MC simulations, is taken as a systematic uncertainty. This variation is found to be between $2\%$ and $16\%$, depending on the rapidity of the $J/\psi$.

A possible contribution from the decay of $Z\to \ell^+\ell^-J/\psi$~\cite{Zpjill1,Zpjill2,Zpjill3} might lead to an enhancement of the measured yields over contributions from \zjpsi.
This possible enhancement is studied by considering the change in the prompt yield after subtracting events for which the mass of the $\ell^+\ell^-J/\psi$ lies within 10~\GeV\ of the world-average value of the $Z$ boson mass; the effect was found to be negligible.

The kinematic acceptance of $Z$ bosons is dependent on the average $Z$ boson polarisation. Due to the high detector acceptance for $Z$ boson decays, the possible effect of modification of the average polarisation of the $Z$ boson in associated production relative to inclusive production is considered negligible in this study.

Uncertainties linked with the luminosity measurement and the $Z$ trigger efficiencies cancel in the ratio of $Z\,+\,J/\psi$ to inclusive $Z$ cross-sections. The contributions of all non-negligible systematic uncertainties are summarised in Table~\ref{tab:summary_of_systematics}.

\section{Results} \label{sec:results}

The results of the two-dimensional maximum likelihood fit are shown in Table~\ref{tab:final_results_with_systematics} for the two rapidity regions. 
\begin{table*}[h!t]
\begin{center}
  \caption{Results of the fit with statistical (first) and systematic (second) uncertainties. The total number of background events is measured in the $2.6<m_{\mu\mu}<3.6\,\mathrm{GeV}$ window. 
The last column presents the expected number of pileup events for the prompt and non-prompt component, and their statistical uncertainty.}
  \begin{tabular}{ l  c  c  c  c}
    \hline
     \multirow{2}{*}{Process} & \multirow{2}{*}{$|y_{J/\psi}|<1.0$} & \multirow{2}{*}{$1.0<|y_{J/\psi}|<2.1$} & \multicolumn{2}{c}{Total}\\ 
	& & & Events found & From pileup \\
\hline
   \\ [-2.0ex]
    Prompt signal 			& $24\pm \ 6  \pm 2$ 	  & $32\pm \ 8 \pm 5$   	& $\ \,\, 56\pm 10\pm 5$
& $5.2 ^{+1.8}_{-1.3}$\\ 
    Non-prompt signal	 	& $54\pm \ 9  \pm 3$	  	  & $41\pm \ 8 \pm 7$    	& $\ \,\,95\pm 12\pm 8$
& $2.7^{+0.9}_{-0.6}$\\
    Background 				& $61\pm 11 \pm 6$		  & $77\pm 13  \pm 7$  	& $138\pm 17\pm 9 $ & \\
   \\ [-2.0ex]
\hline
  \end{tabular}
  \label{tab:final_results_with_systematics}
\end{center}
\end{table*}
The signal significances for both the prompt and non-prompt final states were calculated by performing pseudo-experiments and taking into account the pileup background contribution. 
Events were generated with a di-muon invariant mass and a pseudo-proper time according to the background-only hypothesis, then fitted with the background-only and signal+background hypotheses, 
which allowed the likelihood ratio of the two hypotheses to be calculated and compared with the likelihood ratio of the data.
Using this method, the background-only hypothesis for both the prompt and non-prompt final states was excluded at $5\,\sigma$ significance. To allow for an assessment of the significance beyond
that possible using pseudoexperiments, the significance was extracted as $\sqrt{-2 \times \ln \mathcal{L}}$, where $\mathcal{L}$ is the likelihood ratio of the background-only and signal plus background hypotheses. Both methods yielded consistent results,
the outcome being that
the background-only hypothesis is excluded at $5\,\sigma$ significance for the $Z\,+\,\mathrm{prompt}\ J/\psi$ final state, 
and $9\,\sigma$ significance for the non-prompt $J/\psi$ signature.

After background subtraction, significant signals for the associated-production of $Z\,+$ prompt $J/\psi$ and $Z\,+$ non-prompt $J/\psi$ are observed. 
The background-subtracted $Z\,+$ prompt $J/\psi$ and \znpjpsi candidate yields
are corrected for detector efficiency effects, and production
cross-sections are determined in a restricted fiducial volume given by the
criteria in Table~\ref{tab:phasespace}.
The measured \zjpsi cross-sections are normalised by the inclusive $Z$ production cross-section determined in the same $Z$ boson fiducial volume as the \zjpsi measurement,
benefiting from the cancellation of some systematic uncertainties to allow a more precise determination of production cross-sections. 

\subsection{Fiducial cross-section ratio measurements}

\begin{figure*}[htbp] 
  \centering
		 \includegraphics[width=0.99\columnwidth]{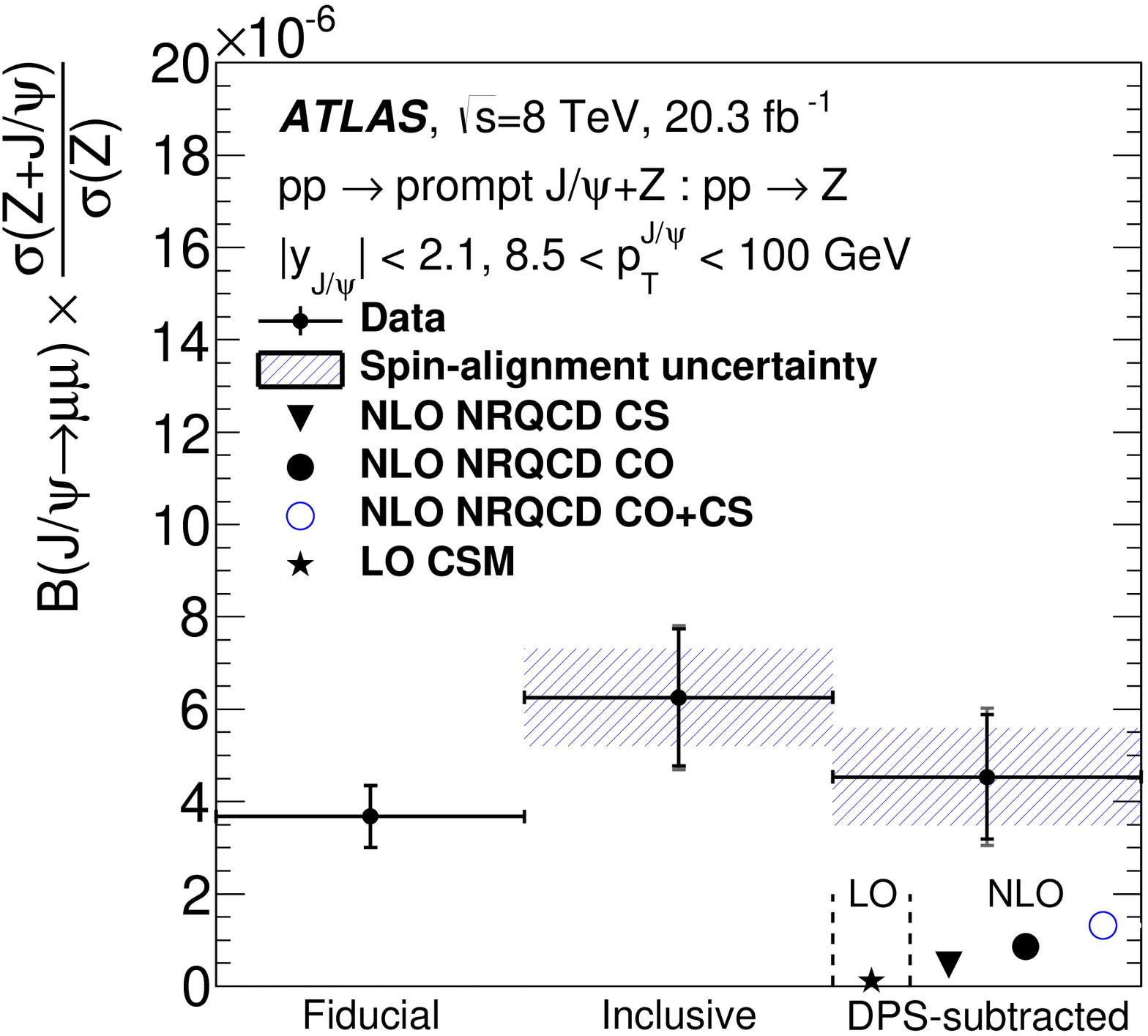}
	          \label{fig:mainxsecresult_prompt}
		 \includegraphics[width=0.99\columnwidth]{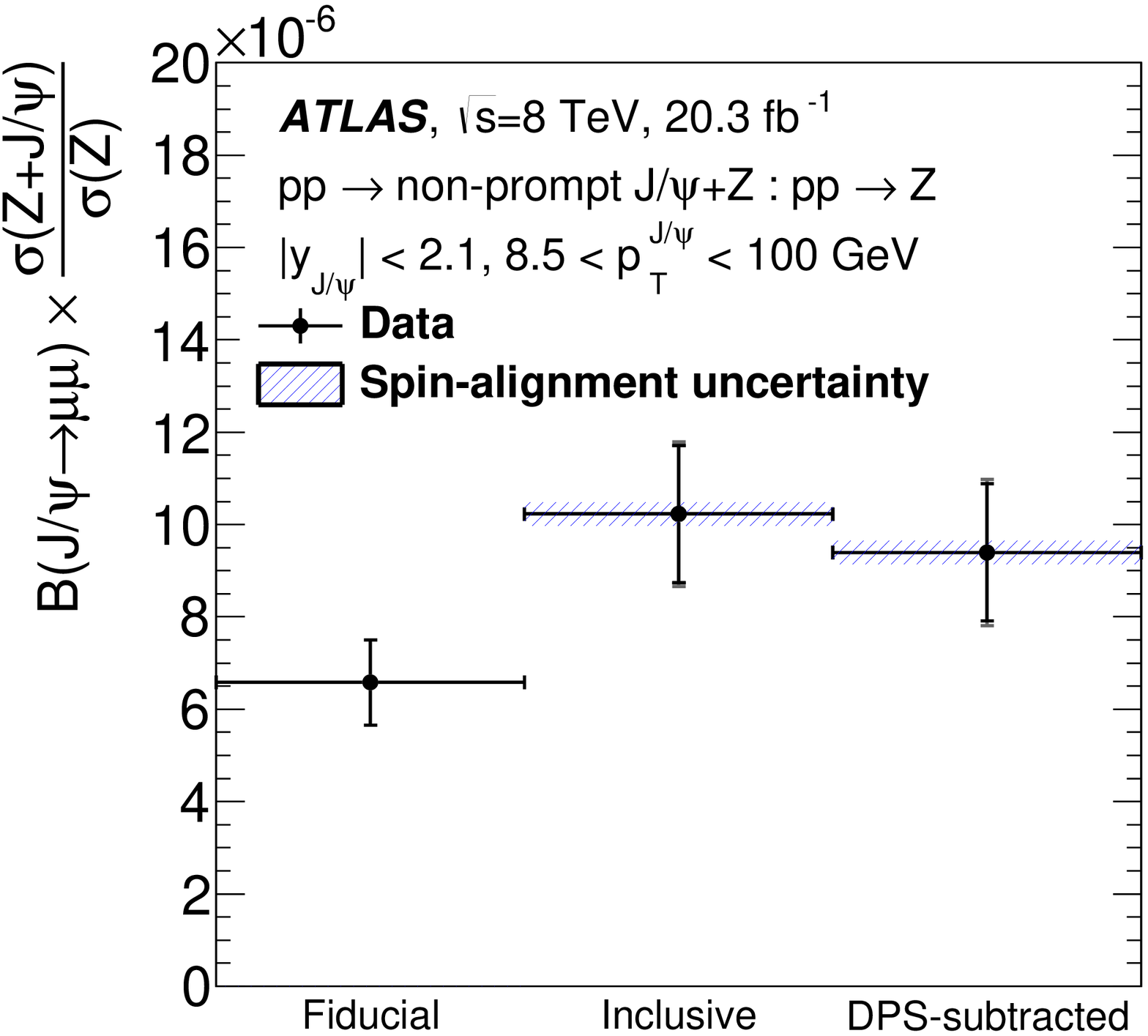}
                	\label{fig:mainxsecresult_nonprompt}
  \caption{Production cross-sections ratios of $J/\psi$ in association with a $Z$ boson, relative to inclusive $Z$ production, for prompt 
and non-prompt $J/\psi$ production. 
    The first point indicates the total integrated cross-section ratio measured in the defined fiducial volume, the second point shows the same quantity corrected for detector acceptance
    effects on the \Jpsi reconstruction, and the third point illustrates the corrected cross-section ratio after subtraction of the double parton scattering contribution as discussed in the text.
    The inner error bars represent statistical uncertainties and the outer error bars represent statistical and systematic uncertainties added in quadrature.
    Also shown are LO~\cite{Gong} and NLO~\cite{ass2} predictions for the inclusive SPS production rates in the colour-singlet (CS) and colour-octet (CO) formalisms.
  }
        \label{fig:mainxsecresult}
\end{figure*}

The fiducial cross-section ratio, as described in Table~\ref{tab:phasespace} (normalised to the inclusive $Z$ boson cross-section), $R^\mathrm{fid}_{\zjpsi}$, is measured without applying corrections for the incomplete geometric acceptance for the $J/\psi$ decay muons, nor for the $Z$ boson acceptance and is defined as\footnote{The equation used is slightly different to that used in the $W\,+\,J/\psi$ analysis~\cite{Aad:2014rua}, which was normalised to unit rapidity.}:

\begin{equation*} 
\begin{split}
R^\mathrm{fid}_{\zjpsi}&=\mathcal{B}(J/\psi\to\mu^+\mu^-)\,\frac{\sigma_\mathrm{fid}(pp\to Z+J/\psi)}{\sigma_\mathrm{fid}(pp\to Z)}\\
&=\frac{1}{N(Z)}\sum_{p_\mathrm{T}\ \mathrm{bins}}\left[N^\mathrm{ec}(Z+J/\psi)- N^\mathrm{ec}_\mathrm{pileup}\right], 
\end{split}
\end{equation*}
\noindent
where $\mathcal{B}(J/\psi\to\mu^+\mu^-)$ is the branching ratio for the decay $J/\psi\to\mu^+\mu^-$~\cite{Beringer:1900zz}, $N^\mathrm{ec}(Z\,+\,J/\psi)$ is the yield of $Z\,+$ (prompt/non-prompt) $J/\psi$ events after corrections for \Jpsi muon reconstruction efficiency,
$N(Z)$ is the background-subtracted yield of inclusive $Z$ events and $N^\mathrm{ec}_\mathrm{pileup}$ is the efficiency-corrected expected pileup background contribution in the fiducial $J/\psi$ acceptance. For prompt and non-prompt production, the cross-section ratios were measured to be:

\begin{equation*} 
  \begin{aligned}
    \textrm{prompt:\ }^\mathrm{p}R^\mathrm{fid}_{\zjpsi}\ &=\ (36.8\pm 6.7\pm 2.5)\, \times\, 10^{-7}\\
    \textrm{non-prompt:\ }^\mathrm{np}R^\mathrm{fid}_{\zjpsi}\ &=\ (65.8\pm 9.2\pm 4.2)\, \times\, 10^{-7}\\
\end{aligned}
\end{equation*}

\noindent
for $8.5\,\mathrm{GeV}<p^{J/\psi}_\mathrm{T}<100\,\mathrm{GeV}$ and $|y_{J/\psi}|<2.1$,
where the first uncertainty is statistical and the second is systematic in origin. The results are summarised in Fig.~\ref{fig:mainxsecresult}.
Production of a $\Jpsi\to\mu^+\mu^-$ meson in association with a $Z$ boson occurs approximately ten times per million $Z$ bosons produced in the fiducial volume defined in Table~\ref{tab:phasespace}.

The differential fiducial cross-section ratios $\mathrm{d}R^\mathrm{fid}_{Z+J/\psi}/\mathrm{d}y$ for prompt and non-prompt \zjpsi production are also determined in two bins, for central \Jpsi rapidities ($|y_{J/\psi}|<1$) and forward \Jpsi rapidities ($1<|y_{J/\psi}|<2.1$), and are reported in Table~\ref{tab:differential_cross_section_results_dy}.

\begin{table*}[htbp]
\begin{center}
 \caption{The fiducial, inclusive (SPS+DPS) and DPS-subtracted differential cross-section ratio $\mathrm{d}R_{Z+J/\psi}/\mathrm{d}y$ as a function of $y_{J/\psi}$ for prompt and non-prompt $J/\psi$.}
 \begin{tabular}{ c c c c}
   \hline
   \multicolumn{4}{c}{Prompt cross-section ratio} \\
   \multirow{2}{*}{$y_{J/\psi}$} & Fiducial $[\times10^{-7}]$ & Inclusive $[\times10^{-7}]$ & DPS-subtracted  $[\times10^{-7}]$\\
    & value $\pm\ \mathrm{(stat)} \pm\ \mathrm{(syst)}$ & value $\pm\ \mathrm{(stat)} \pm\ \mathrm{(syst)} \pm\  \mathrm{(spin)}$ & value $\pm\ \mathrm{(stat)} \pm\ \mathrm{(syst)} \pm\  \mathrm{(spin)}$
\\
   \hline 
   \\ [-2.0ex]
        {$\begin{aligned}[t]
                &|y_{J/\psi}|<1.0 \\
                1.0<&|y_{J/\psi}|<2.1 \\
        \end{aligned}$}
&
        {$\begin{aligned}[t]
   7.6&\pm\ 2.1  &\pm\ 0&.5            \\
   9.8  &\pm\ 2.2  &\pm\ 1&.3         \\
        \end{aligned}$}
&
        {$\begin{aligned}[t]
   13.9&\pm\ 4.6  &\pm\ 0&.8  &\pm\ 3&.4          \\
   15.8  &\pm\ 4.5  &\pm\ 2&.1  &\pm\ 3&.5       \\
        \end{aligned}$}
&
        {$\begin{aligned}[t]
   9.4&\pm\ 4.6  &\pm\ 1&.1  &\pm\ 3&.4          \\
   12.0  &\pm\ 4.5  &\pm\ 2&.7  &\pm\ 3&.5       \\
        \end{aligned}$}
\\
\hline
   \multicolumn{4}{c}{Non-prompt cross-section ratio} \\
   \multirow{2}{*}{$y_{J/\psi}$} & Fiducial $[\times10^{-7}]$ & Inclusive $[\times10^{-7}]$ & DPS-subtracted  $[\times10^{-7}]$\\
    & value $\pm\ \mathrm{(stat)} \pm\ \mathrm{(syst)}$ & value $\pm\ \mathrm{(stat)} \pm\ \mathrm{(syst)} \pm\  \mathrm{(spin)}$ & value $\pm\ \mathrm{(stat)} \pm\ \mathrm{(syst)} \pm\  \mathrm{(spin)}$
\\
   \hline 
   \\ [-2.0ex]
        {$\begin{aligned}[t]
                &|y_{J/\psi}|<1.0 \\
                1.0<&|y_{J/\psi}|<2.1 \\
        \end{aligned}$}
&
        {$\begin{aligned}[t]
   18.0&\pm\ 3.3  &\pm\ 0&.6            \\
   13.5  &\pm\ 2.9  &\pm\ 1&.9         \\
        \end{aligned}$}
&
        {$\begin{aligned}[t]
   29.9&\pm\ 5.0  &\pm\ 0&.9  &\pm\ 1&.1          \\
   19.3  &\pm\ 5.0  &\pm\ 2&.1  &\pm\ 0&.8       \\
        \end{aligned}$}
&
        {$\begin{aligned}[t]
   27.8&\pm\ 5.0  &\pm\ 1&.0  &\pm\ 1&.1          \\
   17.5  &\pm\ 5.0  &\pm\ 2&.1  &\pm\ 0&.8       \\
        \end{aligned}$}
\\

\\ [-2.0ex]
\hline
 \end{tabular}
 \label{tab:differential_cross_section_results_dy}
\end{center}
\end{table*}

\subsection{Inclusive cross-section ratio measurements}

Theoretical predictions for the production rates of \Jpsi are often presented within a limited \Jpsi phase-space, but without any kinematic requirements on the decay products.
To allow comparison of theoretical and experimentally measured production rates, 
corrections derived from simulation are applied to the measured fiducial cross-sections to account for the geometrical acceptance loss due to the muon \pt\ and $\eta$ requirements 
detailed in Table~\ref{tab:phasespace}.
These corrections are dependent on the \pt\ and rapidity of the \Jpsi meson and on the angular distribution of the dilepton system in the decay of prompt \Jpsi. The angular distribution is dependent on the spin-alignment
state of the produced \Jpsi mesons. While the spin-alignment has been measured for inclusive prompt \Jpsi production~\cite{Chatrchyan:2013cla}
and found to be consistent with an isotropic angular distribution hypothesis, \Jpsi produced
in association with a $Z$ boson may have a different polarisation, leading to different decay kinematics. The central value is determined assuming unpolarised decays, with the effect of the most extreme polarisation scenarios assigned as a systematic uncertainty. The largest change in acceptance obtained considering the extreme polarisation scenarios is used as an additional systematic uncertainty in the determination of inclusive production cross-section for prompt $J/\psi$ production, and is equal to $\pm 24\%$ for $|y_{J/\psi}|<1.0$ and $\pm 23\%$ for $1.0<|y_{J/\psi}|<2.1$. The range of variation for non-prompt production was reduced to about $10\%$ of the full range as suggested by the measurement of the $J/\psi$ polarisation in $b$-decays~\cite{Abulencia:2007us} and the uncertainty was found to be $\pm 3\%$ for $|y_{J/\psi}|<1.0$ and $\pm 2\%$ for $1.0<|y_{J/\psi}|<2.1$.

The acceptance-corrected inclusive production cross-section ratio, $R^\mathrm{incl}_{\zjpsi}$, is defined as:

\begin{equation*} 
\begin{split}
R^\mathrm{incl}_{\zjpsi}&=\mathcal{B}(J/\psi\to\mu^+\mu^-)\,\frac{\sigma_\mathrm{incl}(pp\to Z+J/\psi)}{\sigma_\mathrm{incl}(pp\to Z)}\\
&=\frac{1}{N(Z)}\sum_{p_\mathrm{T}\ \mathrm{bins}}\left[N^\mathrm{ec+ac}(Z+J/\psi)- N^\mathrm{ec+ac}_\mathrm{pileup}\right], 
\end{split}
\end{equation*}

\noindent
where $N^\mathrm{ec+ac}(Z\,+\,J/\psi)$ is the yield of $Z\,+$ (prompt/non-prompt) $J/\psi$ events after $J/\psi$ acceptance corrections and efficiency corrections for both muons from the \Jpsi decay, 
$N_\mathrm{pileup}^\mathrm{ec+ac}$ is the expected pileup contribution in the full $J/\psi$ decay phase-space, and other variables are the same as for $R^\mathrm{fid}_{\zjpsi}$. The production cross-section ratio is measured to be:
\begin{equation*} 
  \begin{aligned}
    \textrm{prompt:\ }^\mathrm{p}R^\mathrm{incl}_{\zjpsi}\ &=\ (\ \, 63 \pm 13 \pm 5 \pm 10)\, \times\, 10^{-7}\\
    \textrm{non-prompt:\ }^\mathrm{np}R^\mathrm{incl}_{\zjpsi}\ &=\ (102 \pm 15 \pm 5 \pm\ \,  3)\, \times\, 10^{-7}\\
\end{aligned}
\end{equation*}

\noindent
for $8.5\,\mathrm{GeV}<p^{J/\psi}_\mathrm{T}<100\,\mathrm{GeV}$ and $|y_{J/\psi}|<2.1$,
where the first uncertainty is statistical, the second uncertainty is systematic, and the third uncertainty is due to the unknown $J/\psi$  spin-alignment in $Z\,+\,J/\psi$ production.

The differential fiducial cross-section ratios $\mathrm{d}R^\mathrm{incl}_{Z+J/\psi}/\mathrm{d}y$ for prompt and non-prompt \zjpsi production are also determined in two bins, for central \Jpsi rapidities ($|y_{J/\psi}|<1$) and forward \Jpsi rapidities ($1<|y_{J/\psi}|<2.1$), and are reported in Table~\ref{tab:differential_cross_section_results_dy}.

\subsection{Comparison with theoretical calculations and double parton scattering contributions}

Double parton scattering interactions are expected to contribute significantly to the measured inclusive production cross-sections.
Using the relation in Eq.~\ref{eq:sigmaeff} and a $\sigma_\mathrm{eff}$ value of $15\pm 3\,\mathrm{(stat.)}^{+5}_{-3}\,\mathrm{(syst.)}$~mb, an estimate of the double parton scattering component of the observed signal for both prompt and non-prompt production
can be derived in any kinematic interval of the measurement.
Subtracting this DPS contribution from $R^\mathrm{incl}_{\zjpsi}$ gives an estimate $R^\mathrm{DPS\ sub}_{\zjpsi}$ of the single parton scattering cross-section ratio
for prompt \Jpsi\ production:
\begin{equation*} 
\begin{aligned}
\quad ^\mathrm{p}R^\mathrm{DPS\ sub}_{\zjpsi}\ &=\ (45\pm 13\pm 6\pm 10)\, \times\, 10^{-7}
  \end{aligned}
\end{equation*}
and non-prompt \Jpsi\ production:
\begin{equation*} 
\begin{aligned}
\quad ^\mathrm{np}R^\mathrm{DPS\ sub}_{\zjpsi}\ &=\ (94\pm 15\pm 5 \pm 3)\, \times\, 10^{-7}
\end{aligned}
 \end{equation*}

\noindent
for $8.5\,\mathrm{GeV}<p^{J/\psi}_\mathrm{T}<100\,\mathrm{GeV}$ and $|y_{J/\psi}|<2.1$,
where the first uncertainty is statistical, the second uncertainty is systematic, taking into account uncertainties from the DPS estimate, and the third uncertainty is due to the unknown $J/\psi$ spin-alignment in $Z\,+\,J/\psi$ production. Figure~\ref{fig:mainxsecresult} summarises the fiducial, inclusive and DPS-subtracted cross-section ratios for prompt and non-prompt production
and Table~\ref{tab:differential_cross_section_results_dy} presents the differential cross-section ratios in the central and forward \Jpsi rapidity intervals.
The DPS fraction is $(29\pm 9)\,\%$ for the $Z\,+$ prompt $J/\psi$ signal
and $(8\pm 2)\,\%$ for the non-prompt signal, in the kinematic region studied in this measurement. 

The production cross-section ratios for \zpjpsi production are compared to LO colour-singlet~\cite{Gong} predictions, as well as the contributions 
from colour-singlet (CS) and colour-octet (CO) processes in the non-relativistic QCD (NRQCD) formalism~\cite{ass2}.

All theoretical calculations consider only single parton scattering processes in which the $J/\psi$ mesons are produced directly from the parton interaction, without any feed-down from excited charmonium states. To allow direct comparison to the measured DPS-subtracted cross-section ratios, these predictions are normalised to NNLO calculations of the $Z$ boson fiducial production cross-section ($533.4\,\mathrm{pb}$),
determined using \textsc{fewz}~\cite{Gavin:2010az,Gavin:2012sy}. 

LO colour-singlet mechanism (CSM) predictions for the production cross-section (normalised to the inclusive $Z$ production rate) vary between $(11.6\pm 3.2)\times 10^{-8}$ (from Ref.~\cite{Gong}) 
and $(46.2^{+6.0}_{-6.5})\times 10^{-8}$ (from Ref.~\cite{ass2}). The NLO NRQCD prediction~\cite{ass2} for the colour-singlet rate is $(45.7^{+10.5}_{-\,\, \,9.6})\times 10^{-8}$. 
NRQCD colour-octet contributions to the normalised production rate (that should be added to the corresponding colour-singlet rates to provide the total NRQCD prediction) 
shown in Fig.~\ref{fig:mainxsecresult}
are predicted to be $(25.1^{+3.3}_{-3.5})\times 10^{-8}$ at LO and $(86^{+20}_{-18})\times 10^{-8}$ at NLO accuracy, approximately a factor of two larger
than the contribution from colour-singlet production at the same order in the perturbative expansion.
Uncertainties in the predictions arise from a variation of the renormalisation and factorisation scales up and down by a factor of two from their nominal values, and uncertainties on the charm quark mass.
The variation in the predictions for the colour-singlet rate at LO from different groups arises from a different choice of scale for the central prediction, either taking the $Z$ mass, $m_Z$, or the \Jpsi\ transverse mass, $m_\mathrm{T}^{J/\psi} = \sqrt{ (m^{J/\psi})^2 + (p_\mathrm{T}^{J/\psi})^2 }$, the appropriateness of which is the subject of some discussion~\cite{Gong,ass2}.
The CO predictions presented here use the values for the NRQCD long-distance matrix elements as discussed in Ref.~\cite{ass2}, but do not include uncertainties related to the determination
of these matrix elements\,\cite{Butenschoen:2011yh}.

The effective cross-section regulating multiple parton interactions is expected to be a dynamical quantity dependent on the probed scale of the interactions, and thus should be $x$-dependent (where $x\equiv p_\mathrm{parton} / p_\mathrm{beam}$)~\cite{Abe:1997xk}. Recent theoretical studies~\cite{Blok:2013bpa} have suggested that vector-boson production in association with jets may have $\sigma_\mathrm{eff}$ values as high as $15$--$25\,\mathrm{mb}$. In this paper,  the ATLAS $W\,+\,2$-jet measurement of $\sigma_\mathrm{eff}=15\pm 3\,(\mathrm{stat.}) ^{+5}_{-3}\,\mathrm{(sys.)}\,\mathrm{mb}$ is used to estimate the DPS contribution, and is found to be consistent, within the still sizeable uncertainties, with the observed rates and the plateau observed at small azimuthal separations between the produced $Z$ bosons and \Jpsi, illustrated in Fig.~\ref{fig:deltaphi}. 

The small $\Delta\phi(Z,J/\psi)$ region is sensitive to DPS contributions and can be used to limit the maximum allowed double parton scattering contribution to the observed signal, which corresponds to a lower limit on $\sigma_\mathrm{eff}$, by conservatively assuming that all observed signal in the first bin ($\Delta\phi(Z,J/\psi)<\pi/5$ region) is due to DPS. As the estimated relative signal contribution from DPS processes is largest in prompt production, the data from \zpjpsi\ provides the most stringent limit on the rate of DPS interactions. The data uncertainties and uncertainties inherent in the DPS estimate allow a lower limit $\sigma_\mathrm{eff}>5.3\,\mathrm{mb}\ (3.7\,\mathrm{mb})$ at $68\%\ (95\%)$ confidence level to be extracted from the $Z\,+$ prompt $J/\psi$ data. 

A model-independent upper limit on $\sigma_\mathrm{eff}$ cannot be extracted from these data, as such a limit corresponds to a minimum rate of DPS contribution at small $\Delta\phi(Z,J/\psi)$. While SPS contributions are largest at wide angles, a significant SPS contribution is possible at low angles due to high-order processes~\cite{Lansberg:2013qka}.

\subsection{Differential production cross-section measurements}

\begin{figure}[htbp] 
  \centering
		\includegraphics[width=0.99\columnwidth]{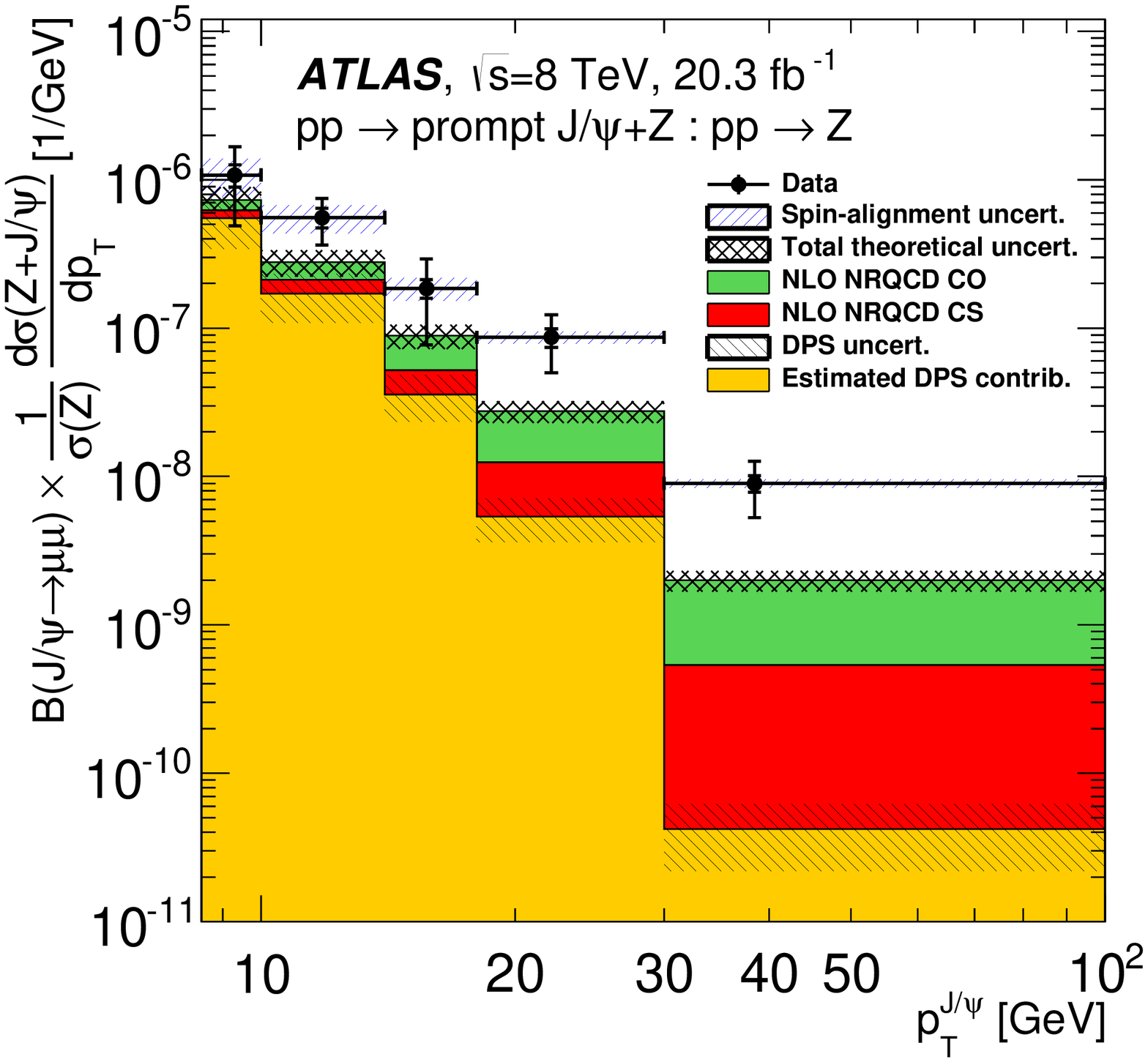}
	          \label{fig:cross_section_prompt}
		\includegraphics[width=0.99\columnwidth]{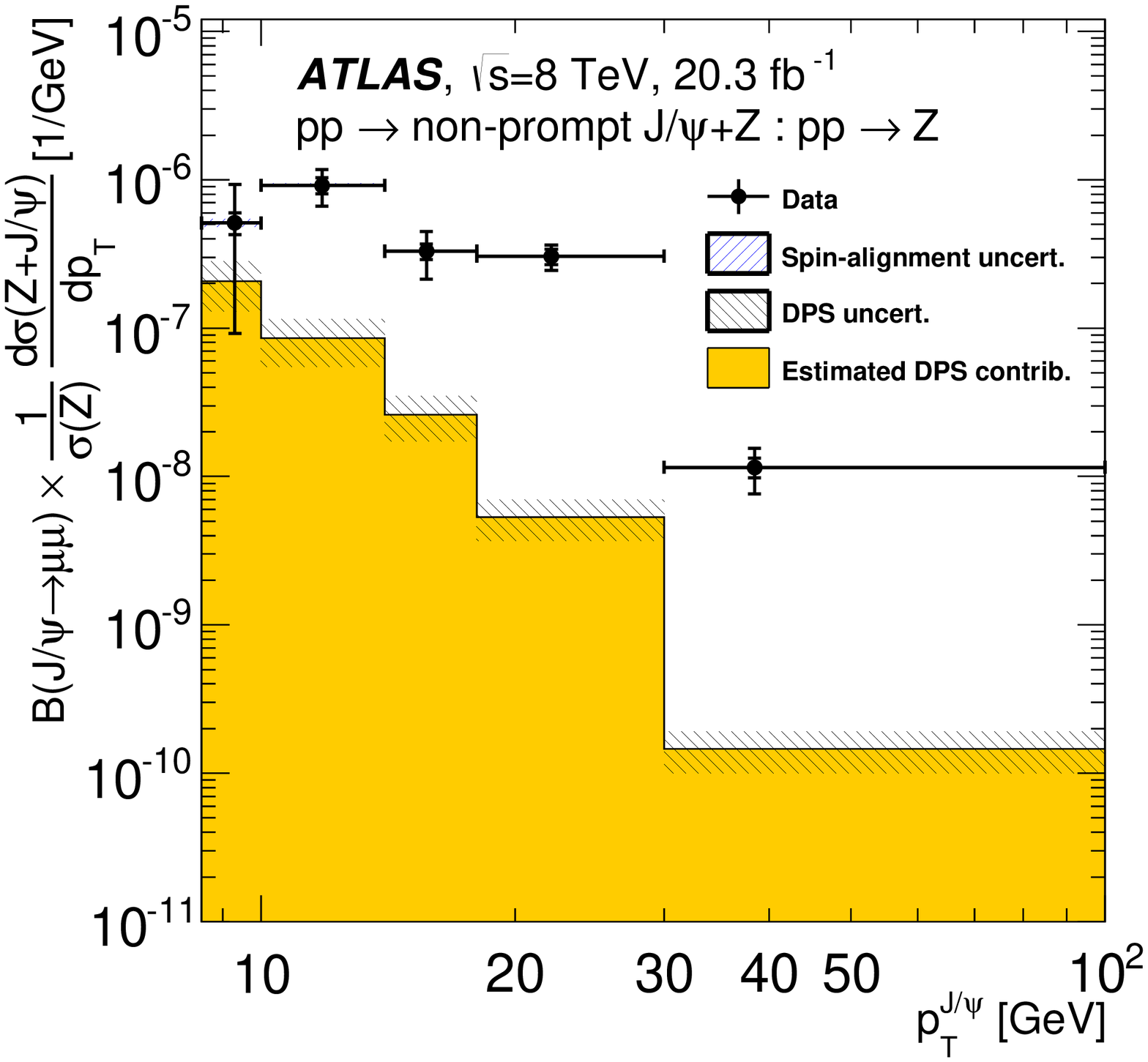}
                	\label{fig:cross_section_nonprompt}
        \caption{Production cross-section of $J/\psi$ in association with a $Z$ boson as a function of the $p_{\rm T}$ of prompt $J/\psi$, and non-prompt $J/\psi$, normalised to the inclusive $Z$ cross-section. Overlaid on the measurement is the contribution to the total signal originating from double parton scattering (DPS) interactions. Theoretical predictions at NLO accuracy for the SPS contributions from colour-singlet (CS) and colour-octet (CO) processes are added to the DPS estimate and presented in comparison to the data as solid bands.
	}
        \label{fig:cross_section}
\end{figure}

Extending upon the measurement of the total inclusive production ratios $R^\mathrm{incl}_{\zjpsi}$ and determination of the DPS contribution, the differential cross-section ratio $\mathrm{d}R_{\zjpsi}^\mathrm{incl}/\mathrm{d}p_\mathrm{T}$ is measured as a function of the transverse momentum of the $J/\psi$ for both the prompt and non-prompt signals, using the sPlot weights obtained from the fit procedure. The differential DPS contribution (using $\sigma_\mathrm{eff}=15\,\mathrm{mb}$) is shown together with the inclusive cross-section ratio in each kinematic interval in Fig.~\ref{fig:cross_section} and in Table~\ref{tab:differential_cross_section_results}. The observed \pt\ dependence is significantly harder than for inclusive \Jpsi production~\cite{JpsiATLAS7TeV}.

\begin{table*}[htbp]
\begin{center}
 \caption{The inclusive (SPS+DPS) cross-section ratio $\mathrm{d}R_{Z+J/\psi}^\mathrm{incl}/\mathrm{d}p_\mathrm{T}$ for prompt and non-prompt $J/\psi$. Estimated DPS contributions for each bin, based on the assumptions made in this study, are presented.}
 \begin{tabular}{ l c c}
   \hline
   \multirow{2}{*}{$p_\mathrm{T}^{J/\psi}\,\mathrm{[GeV]}$} & Inclusive prompt ratio $[\times10^{-7}/\,\mathrm{GeV}]$ & Estimated DPS $[\times10^{-7}/\,\mathrm{GeV}]$\\
    & value $\pm\ \mathrm{(stat)} \pm\ \mathrm{(syst)} \pm\  \mathrm{(spin)}$ & assuming $\sigma_\mathrm{eff}=15\,\mathrm{mb}$\\
   \hline 
   \\ [-2.0ex]
        {$\begin{aligned}[t]
                (8.5&,10) \\
                (10&,14) \\
                (14&,18)\\
                (18&,30) \\
                (30&,100) \\
        \end{aligned}$}
&
        {$\begin{aligned}[t]
   10.8&\pm\ 5.6  &\pm\ 1&.9  &\pm\ 3&.1          \\
   5.6  &\pm\ 1.9  &\pm\ 0&.8  &\pm\ 1&.2       \\
   1.9  &\pm\ 1.1  &\pm\ 0&.1  &\pm\ 0&.3       \\
   0.87  &\pm\ 0.37  &\pm\ 0&.12  &\pm\ 0&.09      \\
   0.090  &\pm\ 0.037  &\pm\ 0&.012  &\pm\ 0&.006 \\
        \end{aligned}$}
&
        {$\begin{aligned}[t]
   5.5 &\pm\  2.1   \\
  1.7  &\pm\   0.6  \\
  0.4  &\pm\  0.1  \\
  0.05  &\pm\  0.02   \\
   0.0004 &\pm\ 0.0002 \\
        \end{aligned}$}
\\\hline
 \multirow{2}{*}{$p_\mathrm{T}^{J/\psi}\,\mathrm{[GeV]}$} & Inclusive non-prompt ratio $[\times10^{-7}/\,\mathrm{GeV}]$\ &
   Estimated DPS $[\times10^{-7}/\,\mathrm{GeV}]$\\
    & value $\pm\ \mathrm{(stat)} \pm\ \mathrm{(syst)} \pm\  \mathrm{(spin)}$ & assuming $\sigma_\mathrm{eff}=15\,\mathrm{mb}$
 \\
   \hline 
   \\ [-2.0ex]
        {$\begin{aligned}[t]
                (8.5&,10) \\
                (10&,14) \\
                (14&,18)\\
                (18&,30) \\
                (30&,100) \\
        \end{aligned}$}
&
        {$\begin{aligned}[t]
   5.1 &\pm\ 4.2  &\pm\ 0&.9 &\pm\ 0&.3 \\
   9.2 &\pm\ 2.5  &\pm\ 1&.2 &\pm\ 0&.3 \\
   3.3 &\pm\ 1.2  &\pm\ 0&.4 &\pm\ 0&.1\\
   3.04 &\pm\ 0.59  &\pm\ 0&.04 &\pm\ 0&.04 \\
   0.115 &\pm\ 0.039  &\pm\ 0&.002 &\pm\ 0&.001\\
        \end{aligned}$}
&

        {$\begin{aligned}[t]
   2.07 &\pm\  0.77   \\
  0.85  &\pm\   0.30  \\
  0.26  &\pm\  0.09  \\
  0.05  &\pm\  0.02   \\
   0.0015 &\pm\ 0.0005 \\
        \end{aligned}$}
\\

\\ [-2.0ex]
\hline
 \end{tabular}
 \label{tab:differential_cross_section_results}
\end{center}
\end{table*}

The measured differential production cross-section ratio for prompt \Jpsi production is compared to NLO colour-singlet and colour-octet predictions. As these predictions are for single parton scattering
rates, the estimated DPS contribution is added to the theoretical predictions to allow like-for-like comparison between theory and data. Theory predicts that colour-octet contributions exceed the production rate from singlet processes
by approximately a factor of two, with colour-octet processes becoming increasingly dominant for higher $p_{\rm T}$ of the \Jpsi. The combination of DPS and NLO NRQCD contributions tends to underestimate the production rate observed in data, with the discrepancy increasing with transverse momentum and reaching a factor of $4$--$5$ at $p^{J/\psi}_\mathrm{T} > 18\,\mathrm{GeV}$. A significant SPS contribution to \znpjpsi\ production rate from $Z+b$-jet production, where the jet contains a \Jpsi meson, is expected but has not been evaluated for this article. The data presented here offer the opportunity to test $Z+b$-jet production at low transverse momentum.

\FloatBarrier

\section{Conclusions} 
\label{sec:conclusions}

This paper documents the first observation and measurement of both associated $Z\,+\ \mathrm{prompt}\ J/\psi$ and \znpjpsi production, with the background-only hypothesis being excluded at $5\,\sigma$ significance for prompt \zjpsi production and at $9\,\sigma$ significance for non-prompt $J/\psi$ production, using $20.3\,\mathrm{fb}^{-1}$ of proton--proton collisions recorded in the ATLAS detector at the LHC, at a centre-of-mass energy of 8~TeV.

Fiducial cross-sections of the production rate of the two final states were measured as ratios to the inclusive $Z$ boson production rate in the same fiducial volume, and found to be
$(36.8\pm 6.7\pm 2.5)\, \times\, 10^{-7}$
and
$(65.8\pm 9.2\pm 4.2)\, \times\, 10^{-7}$
for $Z$ bosons produced in association with a prompt and non-prompt \Jpsi, respectively, where the first uncertainty is statistical and the second is systematic. Ratios, corrected for the limited geometrical acceptance for the muons from the \Jpsi decay in the \Jpsi fiducial volume, are also presented. For prompt production this correction factor depends on the spin-alignment state of \Jpsi produced in association with a $Z$ boson, which may differ from the spin-alignment observed in inclusive \Jpsi production. The measured $Z\,+$ prompt $J/\psi$ production rates are compared to theoretical predictions at LO and NLO for colour-singlet and colour-octet prompt production processes. A higher production rate is predicted through colour-octet transitions than through colour-singlet processes, but the expected production rate from the sum of singlet and octet contributions is lower than the data by a factor of $2$ to $5$ in the $p_\mathrm{T}^{J/\psi}$ range studied.

Measurements of the azimuthal angle between the $Z$ boson and $J/\psi$ meson suggest that both single and double parton scattering contributions may be present in the data. Using the effective cross-section regulating double parton scattering rates as measured by ATLAS in the $W\,+\,2$-jet final state, the fraction of the inclusive production rate arising from double 
parton scattering interactions is estimated to be 
$(29\pm 9)\%$ 
for prompt production and 
$(8\pm 2)\%$
for non-prompt production.
An independent limit on the maximum rate of double parton scattering contributing to the signal is set, corresponding to a lower limit on the effective cross-section of $5.3\,\mathrm{mb}\ (3.7\,\mathrm{mb})$ at $68\%\ (95\%)$ confidence level. The measured production cross-section ratios of inclusive \zpjpsi and \znpjpsi production, and the estimated contribution from double parton scattering, are shown differentially in five intervals of the \Jpsi \pt, with the differential production rates compared to NLO predictions from colour-singlet and colour-octet processes.


We thank CERN for the very successful operation of the LHC, as well as the
support staff from our institutions without whom ATLAS could not be
operated efficiently.

We acknowledge the support of ANPCyT, Argentina; YerPhI, Armenia; ARC,
Australia; BMWFW and FWF, Austria; ANAS, Azerbaijan; SSTC, Belarus; CNPq and FAPESP,
Brazil; NSERC, NRC and CFI, Canada; CERN; CONICYT, Chile; CAS, MOST and NSFC,
China; COLCIENCIAS, Colombia; MSMT CR, MPO CR and VSC CR, Czech Republic;
DNRF, DNSRC and Lundbeck Foundation, Denmark; EPLANET, ERC and NSRF, European Union;
IN2P3-CNRS, CEA-DSM/IRFU, France; GNSF, Georgia; BMBF, DFG, HGF, MPG and AvH
Foundation, Germany; GSRT and NSRF, Greece; RGC, Hong Kong SAR, China; ISF, MINERVA, GIF, I-CORE and Benoziyo Center, Israel; INFN, Italy; MEXT and JSPS, Japan; CNRST, Morocco; FOM and NWO, Netherlands; BRF and RCN, Norway; MNiSW and NCN, Poland; GRICES and FCT, Portugal; MNE/IFA, Romania; MES of Russia and NRC KI, Russian Federation; JINR; MSTD,
Serbia; MSSR, Slovakia; ARRS and MIZ\v{S}, Slovenia; DST/NRF, South Africa;
MINECO, Spain; SRC and Wallenberg Foundation, Sweden; SER, SNSF and Cantons of
Bern and Geneva, Switzerland; NSC, Taiwan; TAEK, Turkey; STFC, the Royal
Society and Leverhulme Trust, United Kingdom; DOE and NSF, United States of
America.

The crucial computing support from all WLCG partners is acknowledged
gratefully, in particular from CERN and the ATLAS Tier-1 facilities at
TRIUMF (Canada), NDGF (Denmark, Norway, Sweden), CC-IN2P3 (France),
KIT/GridKA (Germany), INFN-CNAF (Italy), NL-T1 (Netherlands), PIC (Spain),
ASGC (Taiwan), RAL (UK) and BNL (USA) and in the Tier-2 facilities
worldwide.

\bibliographystyle{atlas4epjc}
\bibliography{zjpsiREFS}

\clearpage\newpage
\restoregeometry\normalsize
\onecolumn
\begin{flushleft}
{\Large The ATLAS Collaboration}

\bigskip

G.~Aad$^{\rm 85}$,
B.~Abbott$^{\rm 113}$,
J.~Abdallah$^{\rm 152}$,
S.~Abdel~Khalek$^{\rm 117}$,
O.~Abdinov$^{\rm 11}$,
R.~Aben$^{\rm 107}$,
B.~Abi$^{\rm 114}$,
M.~Abolins$^{\rm 90}$,
O.S.~AbouZeid$^{\rm 159}$,
H.~Abramowicz$^{\rm 154}$
% ...
\bigskip
\\
% ...
$^{11}$ Institute of Physics, Azerbaijan Academy of Sciences, Baku, Azerbaijan\\
% ...
$^{85}$ CPPM, Aix-Marseille Universit{\'e} and CNRS/IN2P3, Marseille, France\\
% ...
$^{90}$ Department of Physics and Astronomy, Michigan State University, East Lansing MI, United States of America\\
% ...
$^{107}$ Nikhef National Institute for Subatomic Physics and University of Amsterdam, Amsterdam, Netherlands\\
% ...
$^{113}$ Homer L. Dodge Department of Physics and Astronomy, University of Oklahoma, Norman OK, United States of America\\
$^{114}$ Department of Physics, Oklahoma State University, Stillwater OK, United States of America\\
% ...
$^{117}$ LAL, Universit{\'e} Paris-Sud and CNRS/IN2P3, Orsay, France\\
% ...
$^{152}$ Institute of Physics, Academia Sinica, Taipei, Taiwan\\
% ...
$^{154}$ Raymond and Beverly Sackler School of Physics and Astronomy, Tel Aviv University, Tel Aviv, Israel\\
% ...
$^{159}$ Department of Physics, University of Toronto, Toronto ON, Canada\\
% ...
\end{flushleft}


\end{document}